\documentclass[letterpaper]{article} 
\usepackage{aaai2026}  
\usepackage{times}  
\usepackage{helvet}  
\usepackage{courier}  
\usepackage[hyphens]{url}  
\usepackage{graphicx} 
\usepackage{natbib}  
\usepackage{caption} 
\usepackage{algorithm}
\usepackage{algorithmic}
\usepackage{multirow}
\usepackage{makecell}
\usepackage{booktabs}
\usepackage{xcolor}
\usepackage[utf8]{inputenc}
\usepackage{float}
\usepackage{subcaption}
\usepackage{longtable}
\usepackage{float}

\usepackage{newfloat}
\usepackage{listings}
\DeclareCaptionStyle{ruled}{labelfont=normalfont,labelsep=colon,strut=off} 
\floatstyle{ruled}
\newfloat{listing}{tb}{lst}{}
\floatname{listing}{Listing}
\title{The Failed Migration of Academic Twitter: A Case Study of Precocious Adopters}
\author{
    Xinyu Wang\equalcontrib,
    Sai Koneru\equalcontrib,
    Sarah Rajtmajer
}
\affiliations{
    The Pennsylvania State University\\

    University Park, PA, USA\\
    xzw5184@psu.edu, sdk96@psu.edu, smr48@psu.edu

}

\usepackage{bibentry}

\begin{document}

\maketitle

\begin{abstract}
Following changes in Twitter's ownership in 2022 and subsequent changes to content moderation policies, many in academia looked to move their discourse elsewhere and migration to Mastodon was pursued by some. Our study examines the behavior of a self-organized group of early academic adopters who joined Mastodon following changes in Twitter’s ownership. Utilizing publicly available user account data drawn from a voluntarily curated list of academics, we track the posting activity of these early adopters on Mastodon over a one-year period. We also study follower-followee and interaction relationships to map internal networks, finding that the subset of academics who migrated to Mastodon were well-connected. However, this strong internal connectivity was insufficient to prevent users from returning to Twitter/X. Our analyses show that early adopters struggled to maintain engagement, shaped by Mastodon’s decentralized design and competition from alternatives such as Bluesky and Threads. The migration effort lost momentum after an initial surge, as most early adopters reduced activity or returned to Twitter. Our survival analysis further reveals that retention is strongly linked to diverse cross-server engagement and topic-specific server communities. Users with large pre-existing Twitter presence face significantly higher attrition risk, highlighting the challenge of replicating established social connections in a decentralized ecosystem. By examining the coordinated migration attempt of early adopters, we find that even this highly motivated group faced substantial challenges, suggesting that later or less coordinated efforts would likely encounter even greater barriers. 
\end{abstract}


\section{Introduction}
Social media platforms have become essential outlets for communication, information dissemination, and professional networking. Among these platforms, Twitter\footnote{Note on terminology: Twitter rebranded to X in July 2023, which occurred midway through our data collection. For simplicity, we use the terminology Twitter and tweets throughout the paper.} (now known as ``X'') has gained significant popularity and has been likened to a virtual \emph{town square} for its role in facilitating public discourse \cite{mascaro2012Twitter}. Academics and journalists populate sub-communities on Twitter and use the platform for professional communication \cite{jurkowitz2022twitter,malik2019use}. 
In this work, we use the term \emph{Academic Twitter} not to denote a formal organization, but as a shorthand for a diffuse network of scholars who use the platform for professional discourse. Academics have extensively used Twitter to share research findings, announce academic events such as conferences, and contribute to wider scholarly discourse \cite{veletsianos2012higher, mohammadi2018academic, holmberg2014disciplinary}. However, after Elon Musk's acquisition of the company and controversial changes to content moderation policies, many on so-called Academic Twitter sought alternatives \cite{cava2023drivers}. 

Among academics, Mastodon emerged as a popular alternative, with communities organized around shared interests. Unlike traditional platforms, Mastodon operates on a federated model in which independent servers communicate through a shared protocol, allowing users to join different instances while remaining connected across the network \cite{zignani2018follow}. This structure aims to reduce centralized control over internet services by distributing data and services across multiple nodes, enhancing privacy, security, and resilience against censorship and corporate control \cite{raman2019challenges}. Yet despite these advantages, this decentralized structure also leads to lower user engagement compared to centralized platforms like Twitter, where network effects and a larger, more active user base drive higher engagement levels \cite{raman2019challenges}.

The migration of users between social media platforms has attracted research attention, especially in relation to changes in platform policies, user experiences, the factors shaping user decisions, and the resulting implications for online communities.
Social media movements often occur in response to significant changes in platform policies, ownership, user dissatisfaction, or perceived deficiencies in a platform \cite{xu2014retaining}. Qualitative analyses, including interviews and surveys of users migrating from platforms like Fandom to alternatives such as Tumblr, reveal that policy and value-based aspects are primary catalysts of change \cite{fiesler2020moving}. 
Beyond platform-level triggers, prior studies highlight the role of social dynamics in shaping migration behavior. User migration has been linked to perceived platform attractiveness and peer effects \cite{hou2020understanding}. In the context of Twitter’s ownership change, migration to Mastodon reflected social influence, network structure, commitment, and shared identity, while remaining characterized by low Twitter account deletion rates, concentration of users in large Mastodon instances, and divergent posting behaviors across platforms \cite{cava2023drivers, he2023flocking}. Recent work raises questions about sustainability, reporting lower engagement and posting activity on Mastodon compared to Twitter, and pointing to challenges in retaining active users within a decentralized architecture \cite{jeong2023exploring}. 

Despite prior work characterizing general trends across the population of platform users, there remains a lack of targeted research concerning migration of narrower and highly interconnected communities characterized by shared motivations and coordinated efforts. We address this gap by providing a targeted case study of early academic adopters responding to platform changes, examining their migration patterns and behaviors to offer a more nuanced understanding of how these users navigate platform transitions. Given the significant role that Twitter plays in many academics' professional lives \cite{knight2016tweet,klar2020using}, academics represent a valuable subpopulation for studying migratory dynamics. Our work addresses the following research questions:

\vspace{0.1cm}

\noindent \textbf{RQ1:} 
How did early academic adopters navigate the migration to Mastodon, and what do longitudinal, network, and cross-platform patterns reveal about their retention?

\noindent \textbf{RQ2:} 
Which measurable user and community characteristics are associated with long-term retention on Mastodon?

\vspace{0.1cm}

Our findings indicate that the majority of academic researchers who moved to Mastodon failed to maintain their posting activity and follower engagement over the long term. Although migrated academic accounts formed strongly-connected internal networks, these connections were insufficient to sustain user engagement over time. Of note, however, communities that established field-specific server instances had higher user retention rate compared to those on general-purpose servers. Cross-platform analysis of most recent Twitter activity showed that most users returned to Twitter. Survival analysis confirms that users' persistence is associated with a number of factors including diverse cross-server engagement, early platform activity, choice of server. Furthermore, cross-platform modeling reveals a divergent pattern, with a longer history of prior activity on Twitter associated with higher attrition risk. These results underscore the challenges and opportunities associated with building and maintaining communities on new platforms.

\section{Related Work}

\subsection{Cross-Platform Migration}
User movement between social networking services has been studied as a migration process shaped by contextual factors, including dissatisfaction with platform experience, social influence, and perceived migration costs \cite{xu2014retaining}. Early work analyzed it using Push Pull Mooring framework with Push effects (e.g., policy changes) drive users away, while Pull effects (e.g., better features) attract them to alternatives. However, Mooring factors such as switching costs often prevent complete migration \cite{chang2014push, lin2017understanding}. 
Evidence from user surveys indicates that platform migration, such as movements from Fandom to Tumblr, is often driven by policy changes and misalignment in platform values \cite{fiesler2020moving}. Additionally, research has shown that user migration is influenced by factors such as perceived platform attractiveness and peer influence \cite{hou2020understanding}. Study of users migrating from Twitter to Mastodon post-Elon Musk's acquisition indicated that social influence, network structure, commitment, and shared identity significantly drive collective behavior changes in online communities \cite{cava2023drivers}. Analysis showed that Twitter to Mastodon migration is characterized by low Twitter account deletion rates, concentration of users in large Mastodon instances, varied posting behaviors across platforms \cite{he2023flocking}. A similar study on Twitter-to-Mastodon migration reveals lower user engagement and posting activity on Mastodon, highlighting challenges in retaining active users due to its decentralized architecture \cite{jeong2023exploring}. These results are consistent with evidence that the transition is often incomplete with users maintaining cross-platform presence rather than fully substituting one platform to the other \cite{pena2025straddling}. More broadly, recent work on platform switching from Twitter shows that users disperse across multiple destination platforms rather than relocating to a single successor platform \cite{radivojevic2025user}.

Our work looks more deeply at the specific case of the relatively well-synchronized and coordinated movement of communities of academic users across platforms.

\subsection{Deplatforming as Involuntary Migration}

Unlike voluntary platform migration, which is often motivated by user dissatisfaction or policy changes, deplatforming represents a form of forced migration where users are compelled to leave due to removal or suspension \cite{rogers2020deplatforming}. 
A study of Reddit's 2015 ban shows that deplatforming can reduce the visibility of harmful content on mainstream platforms while simultaneously displacing communities to alternative or less moderated spaces \cite{chandrasekharan2017you}. Parler was deplatformed after the 2021 U.S. Capitol attack, while Gab expanded as a refuge for banned Twitter users; however, research shows that such platforms often attract smaller or more fragmented audiences and face challenges related to moderation and community stability \cite{aliapoulios2021early, zannettou2018gab}. While distinct from voluntary migration, deplatforming likewise underscores how platform-level conditions shape user mobility and contextualizes the broader challenges communities face when establishing presence on alternative platforms. 

\subsection{Characteristics of Mastodon}
The decentralized structure of Mastodon and similar platforms leads to unique user engagement dynamics that are different from centralized platforms.  
Research suggests that users are drawn to Mastodon for the ease of protecting their information from data mining and the absence of algorithms shaping their feeds \cite{lee2023uses}. However, this can also make content discovery more difficult \cite{zignani2019footprints} and may contribute to network fragmentation \cite{kleppmann2024bluesky}, shaping interactions and communication patterns that differ from centralized platforms \cite{brauweiler2025decentralized}. The decentralized structure encourages strong, tight-knit communities within individual server instances but fewer connections across instances \cite{zignani2018follow}. These characteristics pose challenges for migration as well. Our study examines how specialized communities—particularly academics—navigated these dynamics in their movement from Twitter to Mastodon.

\section{Data}\label{sec:Data}

\noindent Our analyses are primarily based on a public GitHub repository containing lists of academics on Mastodon\footnote{\url{https://github.com/nathanlesage/academics-on-mastodon}}. These lists were curated by researchers from 50 distinct disciplines, who voluntarily added their own names and accounts. Each list corresponds to a research discipline, containing account details of researchers within that area, their Mastodon IDs, and a short description of their research. Some lists also included each user's Twitter handle. For the purpose of longitudinal analysis, we focused only on scholars whose Mastodon accounts were already in the repository at the beginning of November 2022, noting that the Twitter-to-Mastodon migration began in late October 2022. That is, the users in our dataset were part of a \emph{coordinated, synchronized migration effort} in November 2022.

In total, we collected the Mastodon profiles and corresponding public metrics, e.g., number of followers, following, and timestamps of most recent status for \textbf{7,542 academic user accounts} using a customized Mastodon API\footnote{\url{https://github.com/halcy/Mastodon.py.git}}. We collected data weekly from November 2022 to October 2023 allowing for temporal analysis of user activity and engagement over the span of a year. In addition, we collected the most recent Mastodon data related to follower and following accounts for each scholar to conduct network analyses based on fields of study and associated servers. 
Additionally, to capture dynamics beyond follower and followee connections, we collected the user interactions (replies, boosts, mentions, and favorites) covering the full observation window from November 2022 to October 2023. We collected activity for users in our sample to extract both outgoing interactions and incoming interactions. Note that this interaction dataset was collected retrospectively in December 2025.

For cross-platform analyses, we used self-reported Twitter handles to gather scholars' profile information. Of 7,542 Mastodon scholars, a subset of 3,131 scholars accurately provided Twitter handles suitable for data collection. Using the Twitter (``X'') Basic API, we retrieved profile information, most recent tweet IDs, and engagement metrics. Additionally, we used the Twitter ID collection endpoint to extract the posting time of their most recent tweet.

Finally, to understand discussions surrounding Twitter migration to Mastodon on both platforms, we collected posts from both platforms during the one year window. Specifically, we collected tweets via Zeeschuimer\footnote{https://github.com/digitalmethodsinitiative/zeeschuimer} using the boolean search ``mastodon (researcher OR researchers OR academic OR academics OR professor OR professors OR scholar OR scholars) lang:en until:2023-11-01 since:2022-10-01''. 
The set of academic-related terms was developed through a preliminary scan of tweets, ensuring coverage of both individual (e.g., “professor,” “scholar”) and collective (e.g., “academics,” “researchers”) identifiers. Since many tweets were included because the display name contained the word ``mastodon'' (as some users use their Mastodon account name as their Twitter display name) we further filtered out these tweets. We retained only those that contained relevant keywords in the main body of the tweet, resulting in a total of 1,497 tweets.
We also collected posts through the Mastodon API using migration-related hashtags adapted from prior work \cite{jeong2024exploring}. We excluded the tag "MastodonSocial" because it is commonly used for general content on Mastodon rather than migration. From a total pool of 12,654 migration related posts, we then applied a second stage filter using the same set of academic keywords described earlier to identify content relevant to academic migration. This process yielded 129 posts in the final dataset.

\section{
RQ1: Migration Dynamics and User Retention}
Following, we present findings from multiple analyses conducted to understand the dynamics of the migration. First, we used the longitudinal data from Mastodon to understand user attrition and engagement trends over the year. Then, we used network analysis to understand the connectivity and distribution of academics across Mastodon servers. Next, we conducted a cross-platform analysis using the subset of users who provided Twitter handles comparing their activity patterns on Twitter and Mastodon. Taken together, these analyses identify patterns and potential factors associated with user retention, which in turn motivate the selection of features included in the survival analysis for RQ2.

\subsection{Longitudinal Patterns of Adoption and Retention}\label{sec:longitudinal}

\subsubsection{Attrition of Academic Accounts on Mastodon}

Figure \ref{fig:combined_dist} (Top) gives a bar plot illustrating the change in number of active users aggregated by month and a line plot showing the monthly retention rate based on the previous month. 
In this context, we considered a user to be active if they made at least a post in a given month.
In the initial month of data collection (November 2022), there were 7,505 active scholars who had posted within that same month.\footnote{1 scholar (April 2022), 4 (May 2022), and 32 (October 2022) had their last activity outside our collection window. We omit them from the bar plot but include them in subsequent analyses.} Over the next seven months, the number of active scholars experienced a decline, showing a minor resurgence in July 2023 before continuing its descent, ultimately reaching 2,398 by October 2023. We observe a substantial rate of attrition; each month, approximately 10\% to 20\% of  scholars in our set stopped actively using the platform. The surge in July 2023 coincided with significant shifts in Twitter's policy on limiting number of posts user can view and officially rebranded to ``X'' which may have prompted a renewed but temporary interest in Mastodon as alternative.

\subsubsection{Engagement with Academic Accounts on Mastodon} 
\begin{figure}[t]
    \centering
    \includegraphics[width=1\linewidth]{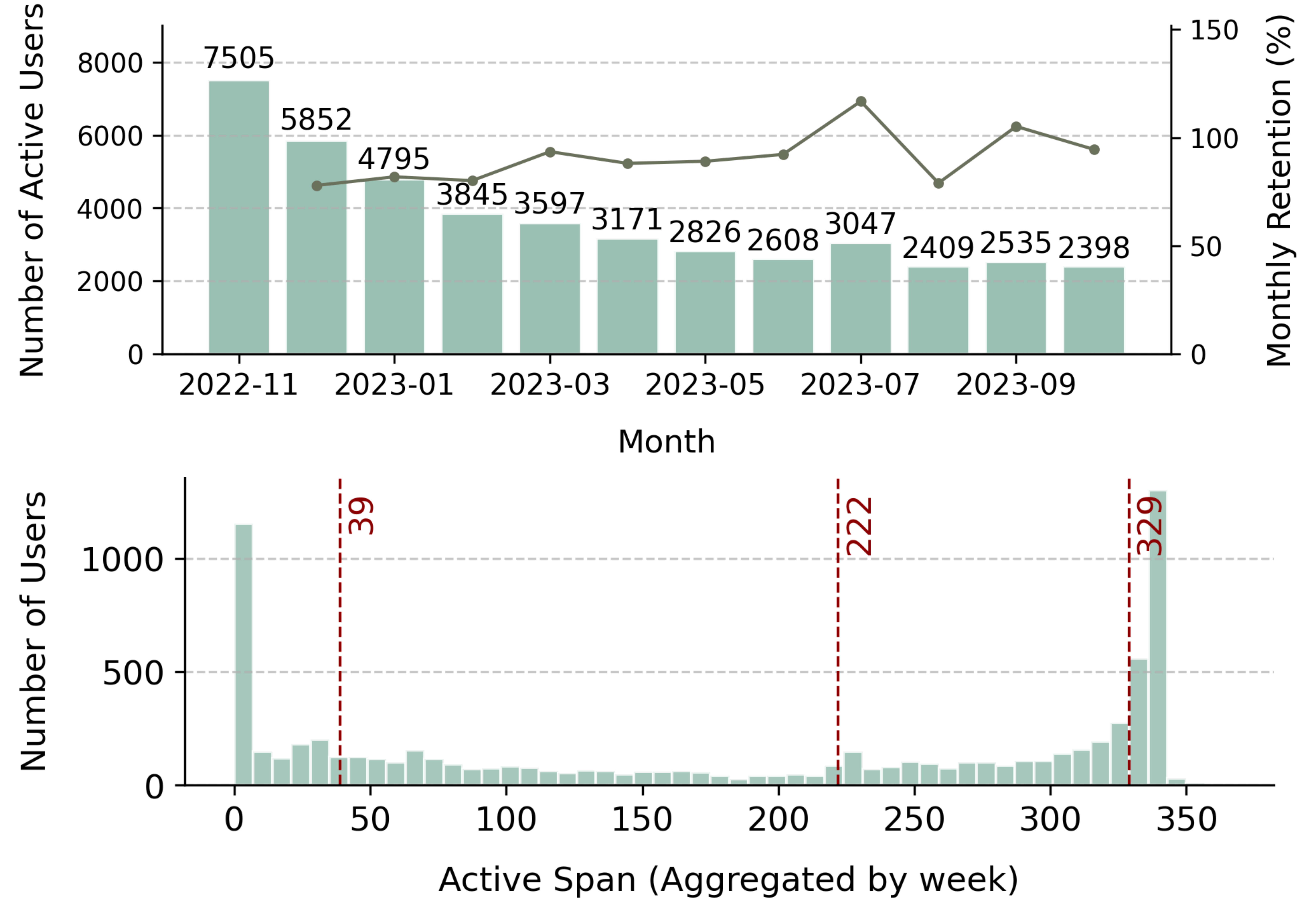}
    \caption{  \textbf{Top}: Histogram showing number of monthly active users (active=at least one post per month). \textbf{Bottom}: Histogram showing distribution of user active spans (number of days they are on Mastodon aggregated by week).}
    \label{fig:combined_dist}
\end{figure}

Given the central role of engagement in user activity, we look at the engagement with migrated academic accounts on Mastodon. Figure~\ref{fig:combined_dist} (Bottom) presents the distribution of user active spans, which we use to characterize accounts into activity patterns for comparison across dimensions in later sections.

\vspace{0.1cm}
\noindent\textit{Active users. }
We define active users as scholars who engaged on Mastodon during the initial Twitter-to-Mastodon migration, which began in late October 2022. For our purposes, we've defined this initial period between October 15 and November 30, 2022. 
We categorize active users into four engagement groups based on the empirical distribution of active spans using the 25th, 50th, and 75th percentiles: one-time visitors; short-term, long-term, and persistent adopters. These data-driven cutoffs reflect the underlying heterogeneity in the duration of their activity on Mastodon following the migration. The days cutoffs are rounded to the nearest integer.

\vspace{0.1cm}
\noindent \textit{One-time visitors.} 
These academics engaged only briefly at the beginning of the Twitter-to-Mastodon migration, with active spans falling below the 25th percentile of the overall distribution ($n_{days}\leq39$). Their minimal participation suggests exploratory behavior during the migration wave without continued investment in the platform.

\vspace{0.1cm}
\noindent \textit{Short-term adopters.} 
This group falls between the 25th and 50th percentiles of active spans ($39<n_{days}\leq222$). These users engaged beyond initial exploration but did not sustain participation long enough to become stable contributors, potentially reflecting unmet expectations or competing commitments on other platforms.

\vspace{0.1cm}
\noindent \textit{Long-term adopters.} 
These users fall between the 50th and 75th percentiles of the distribution ($222<n_{days}\leq329$). They demonstrated extended engagement following the migration but did not remain active through the full analytic period.

\vspace{0.1cm}
\noindent \textit{Persistent adopters.}  
These users fall above the 75th percentile ($n_{days}>329$), corresponding to the longest active spans through the end of our analyses. Their sustained engagement suggests successful integration into Mastodon's academic community and continued participation over time.

\vspace{0.1cm}
\noindent \textit{Inactive users.} Academics in this category were already using Mastodon before the migration wave from Twitter. Despite their early presence on the platform, they remained inactive during the migration period. One user was inactive during the first month but engaged briefly afterward. Inactive users are not included in subsequent analyses. The distribution of users is given in Table~\ref{tab:user_categories}.

\begin{table}[t]
\small
\centering
\begin{tabular}{|l|l|r|}
\hline
\textbf{User Category} & \textbf{Subcategory} & \textbf{Count} \\ \hline
{Active} & One-time visitors &  1895\\
 & Short-term adopters & 1876\\
 & Long-term adopters &  1921\\
 & Persistent adopters &  1845\\ \hline
{Inactive}& One-time visitors & 4\\\
 &Returned Users&1\\
 \hline
\end{tabular}
\caption{Number of users, by activity. }
\label{tab:user_categories}
\end{table}

We calculate the average following, follower, and post counts for each user category, aggregated by month as shown in Figure~\ref{fig:metrics}. Our analysis reveals a sharp increase in all three public metrics at the onset of the migration period, specifically from November 2022 to December 2022. However, beyond this period, the rate of increase for following and follower metrics slows considerably. For one-time, short-term, and long-term users, growth in following and follower counts stagnated beginning in early 2023. As these users were not regularly contributing content, their visibility and potential to attract followers were likewise limited. In contrast, persistent users continued to post regularly. Despite their consistent activity, however, growth in their connections remained slow. This slow growth highlights the challenges faced in establishing and expanding social ties and communities on Mastodon. The persistent users' continued efforts to engage and post suggest a commitment to the platform, yet the limited increase in followers indicates potential barriers in user retention and community building. 

\subsubsection{Field-specific Activity Differences among Academic Accounts on Mastodon}

As discussed in Section Data, users self-identified their research disciplines in public lists. We aggregate these disciplines into 9 broad fields, consistent with existing taxonomies of academic disciplines \cite{bepress2019disciplines}. We adopt this taxonomy because its granular disciplinary structure aligns more closely with the self-reported research areas in our dataset, and it is widely deployed in institutional repository metadata systems (e.g., \footnote{https://guides.lib.odu.edu/c.php?g=502940\&p=5839940},\footnote{https://stars.library.ucf.edu/faq.html},\footnote{https://digitalcommons.lib.uconn.edu/about.html}). Scholars identifying with multiple disciplines and open science researchers were categorized as multi-disciplinary researchers. Inactive users were excluded from subsequent analyses. Disciplinary statistics of the 7,537 academics in our dataset are provided in Table~\ref{tab:field_statistics} in the Appendix.

We identify the proportion of one-time, short-term, long-term, and persistent users by field (see Table~\ref{tab:percentage} in the Appendix). The highest proportion of persistent users is observed among multidisciplinary researchers. One possible explanation is that interdisciplinary scholars engage with discussions spanning multiple fields, which may provide a broader set of conversational opportunities and connections that support continued participation. This observation motivates the inclusion of a binary indicator for interdisciplinary research in our subsequent modeling of user retention.

\subsection{Network Structure and Community Formation}

To understand the connectivity within and between different fields, we collected the complete follower-followee network and multiple types of interaction networks of all academic accounts on Mastodon, and constructed an internal connection network within each academic group, as well as a network aggregated by field of study. 
\subsubsection{Internal Networks Aggregated by Field}
\begin{figure}[ht]
    \centering
    \includegraphics[width=\linewidth]{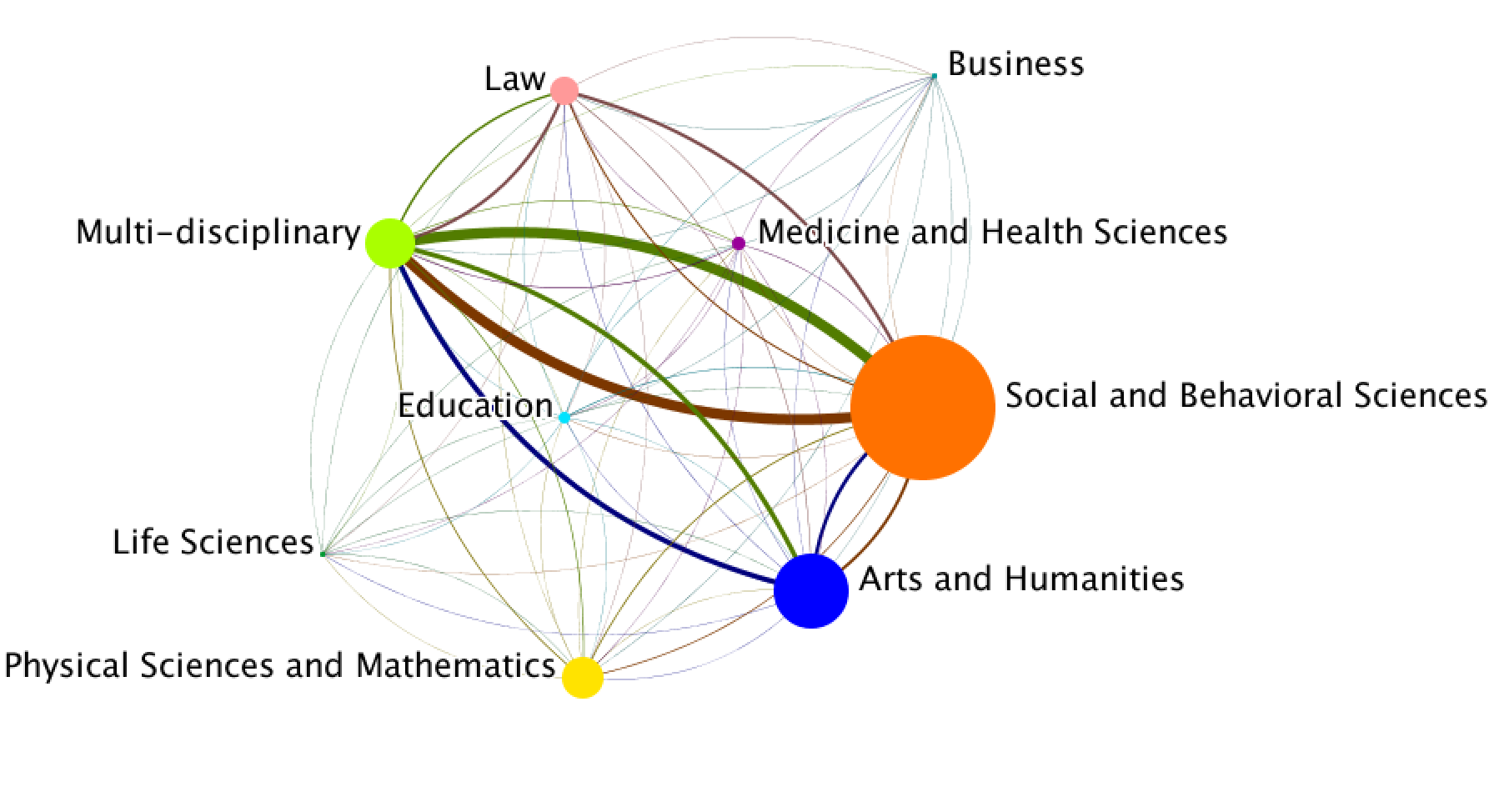}
    \caption{Internal field network (size: number of scholars in each field; edge weights: frequency of connections; edge color: same as the source; self-loops are omitted for better visualization.)}
    \label{fig:field}
\end{figure}

Figure~\ref{fig:field} shows the network connections across different fields. In this network, node size represents the number of academics in each field, while edge width indicates the frequency of connections between the user accounts from different fields. Given the nature of the field, certain fields such as \textit{Social and Behavioral Sciences} are strongly connected to \textit{Multi-disciplinary} field. Our analysis reveals that the academic migration has led to the formation of a well-connected internal network, suggesting that researchers actively attempted to establish a new community on Mastodon. 
However, despite these coordinated efforts, the migration still faced significant challenges, as reflected through the low retention rates reported in the previous section.

\begin{figure}[h]
    \centering
    \includegraphics[width=0.95\linewidth]{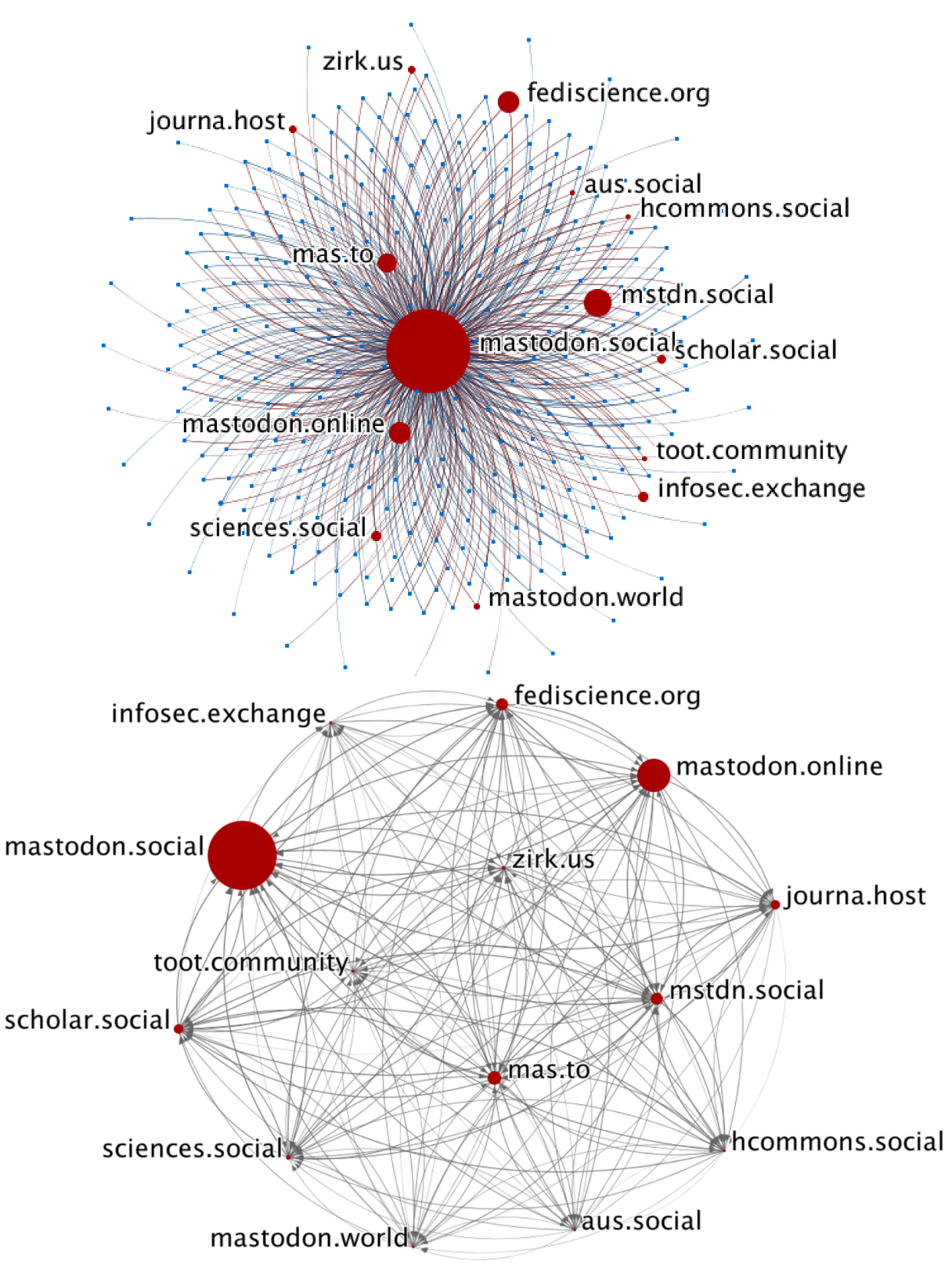}
\caption{(\textbf{Top}) Server network constructed from internal follower–followee relationships. Isolated servers are removed for clarity. Node size is scaled by the number of scholars within each server. Servers with more than 100 scholars are highlighted in red. (\textbf{Bottom}) Internal comment interaction network among servers with more than 100 scholars. Node size is scaled by the total number of interactions for each server, and edge width is scaled by the number of interactions between servers. Internal boost, favorite, and mention networks are reported in Figure~\ref{fig:rest}.} 
    \label{fig:server}
\end{figure}

\subsubsection{Internal Networks Aggregated by Server}
Prior studies have shown that the Mastodon's server instance-based structure makes it easier for users to discover and connect with others within the same instance than outside the instance. This leads to stronger within instance ties compared to weaker links across instances \cite{zignani2019footprints}. Therefore, we analyzed the follower-followee network aggregated by Mastodon instances and the corresponding interaction networks among top servers, shown in Figure~\ref{fig:server}. 

The academic users were spread across a total of 342 servers, with 9 server instances being isolates without any connecting edges to other instances, which were removed for clarity. Table \ref{tab:server_centrality_metrics} in the Appendix shows network centrality metrics for each server and the distribution of the users across different servers. It also includes the official description of the server instances, indicating whether they are customized to a specific group of scholars. During the migration, different disciplines used different strategies: some used general-purpose servers like \textit{mastodon.social} which host a wide variety of content beyond just academic content, while others took a coordinated approach on setting up their field-specific server instances such as \textit{infosec.exchange} and \textit{journa.host}. Figure \ref{fig:server} (Top) shows the follower-followee connections across different servers showing certain general purpose instances such as \textit{mastodon.social} are popular among the tracked academic users (containing 21.3\% of the total) than other general purpose servers like \textit{mas.to}. We observe a central hub pattern in which the majority of follower–followee connections involve the mastodon.social server and other servers, forming a radiating structure. To examine interactions beyond follower-followee relationships, we further plot interaction networks among top servers with more than 100 scholars (highlighted in red in Figure~\ref{fig:server} (Top)). Figure~\ref{fig:server} (Bottom) shows the internal comment interaction network (boost, favorite, and mention networks are presented in Figure~\ref{fig:rest} in the Appendix). Node size is scaled by the total number of interactions for each server, and edge width is scaled by the number of interactions between server pairs.
We find that even when servers are not directly connected through follower–followee relationships, they are connected through interactions such as comments, mentions, and boosts. This motivates the inclusion of both within-server and inter-server interaction in subsequent modeling, as these interaction-based connections capture federated engagement that is not visible in the follower–followee network alone. In particular, interaction networks reveal cross-server communication pathways that operate independently of stable follower ties.

\begin{figure}[ht]
    \centering
    \includegraphics[width=1\linewidth]{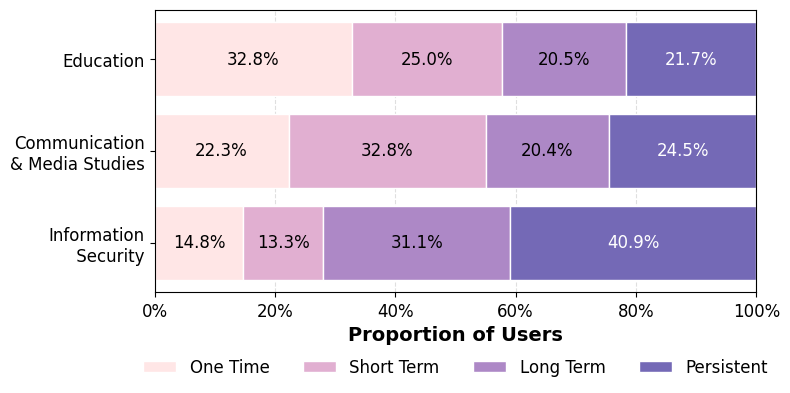}
    \caption{User retention breakdown for Information Security scholars compared to the sub-fields that are similar in size (Info. Sec.: n=264; Comm. \& Media Studies: n=265; Education: n=244.)}
    \label{fig:case_bar}
\end{figure}

\noindent \textit{Comparison Across Communities.}
To explore how a discipline specific server could influence user retention and engagement, we isolated the internal follower-followee network of accounts within \textit{Information Security}. Figure~\ref{fig:infosec} shows the internal network of the discipline, where most scholars congregated on a single server: \textit{infosec.exchange} (highlighted in red). We compare the retention roles of scholars from the discipline with those from a discipline where users were concentrated in a general-purpose server and other sub-fields of similar size (Education and Communication \& Media Studies), as shown in Figure~\ref{fig:case_bar}. We see a significantly higher percentage of persistent users amongst information security scholars (40.9\%) in comparison to other comparable disciplines. This suggests that establishing field- or discipline-specific communities or servers could foster a stronger sense of belonging, which may help retain scholars who have migrated, reducing the likelihood of their return to Twitter or seeking alternative platforms.

\subsection{Cross-Platform Patterns and Usage Shifts}

We conducted a cross-platform analysis to analyze the state of the early adopters' accounts on Twitter. Approximately 50\% of the scholars voluntarily provided their Twitter usernames, which were accurate and traceable to their Twitter profile pages. Using the Official Twitter Basic API, we collected the profile information of these accounts. Despite some account deletions, we successfully retrieved 3,131 Twitter accounts that could be matched to Mastodon accounts. Next, we retrieved the most recent posting times of these Twitter accounts to assess their activity status. 
\begin{table}[b]
\centering
\scriptsize
\begin{tabular}{c|c|c|c}
\toprule
&Count&\textbf{Following Count}&\textbf{Follower Count}\\
\midrule
\textbf{Active}&2496&$\mu$:1989.07, $\sigma$:3087.68&$\mu$:13208.76 , $\sigma$:108495.54 \\
\textbf{Semi-active}&220&$\mu$:1137.56 , $\sigma$:1263.87 &$\mu$:4467.28, $\sigma$:19152.90  \\
\textbf{Inactive}&91&$\mu$:814.56 , $\sigma$:1062.39 &$\mu$:2401.46 , $\sigma$:5203.50  \\

\textbf{Locked}&324&$\mu$:1859.72 , $\sigma$:2300.91&$\mu$:6347.10 , $\sigma$:31583.98  \\
\bottomrule
\end{tabular}
\caption{Engagement statistics of 3,131 scholars (number of users per category with mean and standard deviation of following and follower counts).}
\label{tab:user_distribution}
\end{table}

Based on their posting behavior, we categorized the Twitter account activity into the following statuses:

\vspace{0.1cm}
\noindent \textit{Locked.} Scholar's account is locked down, meaning their posts are private and cannot be accessed.

\vspace{0.1cm}
\noindent \textit{Inactive.} Scholar's latest tweet was posted before December 31, 2022, and their posts were made before this date, indicating no recent activity since the start of the year.

\vspace{0.1cm}
\noindent \textit{Semi-active.} Scholar's latest tweet was posted between December 31st, 2022 and November 1, 2023. This indicates some activity during the year but no recent posts after November, 2023, the end of our Mastodon tracking.

\vspace{0.1cm}
\noindent \textit{Active.} Scholar's latest tweet was posted after November 1st, 2023, indicating continuous activity and engagement on the platform.

\vspace{0.1cm}
Using the above categorization, we calculated the count and connection statistics for each category, and all statistics are presented in Table~\ref{tab:user_distribution}. We observe that 79.72\% of scholars are still active on Twitter, indicating no signs of migration or intention to fully migrate. These active Twitter users have accumulated approximately 10 times more  followers, which makes it more difficult to give up the platform.
Some active users still make their Mastodon information accessible to the community. For instance, one scholar noted:
\begin{quote}
\textit{``Hi ya’ll for those moving over to \#Mastodon you can currently find me $<$Scholar's Mastodon Account$>$."}
\end{quote}
This group of scholars does not personally intend to migrate but aims to maintain connections with those who do intend to fully migrate to Mastodon. 
Furthermore, 7.03\% of the scholars are semi-active users on Twitter. They did not stop using Twitter during the migration, but have reduced their presence gradually during a one-year grace period, potentially transitioning away from Twitter entirely.  In contrast, 2.91\% of the scholars still have accessible but inactive accounts since November 2022, indicating they will not return to Twitter. For example, one scholar's last post reads,
\begin{quote}
\textit{``I’m leaving Twitter but I’ve been advised not to delete my account in order to avoid possible misuse - so I’ll leave it here as a placeholder. See you at Mastodon!"}
\end{quote}

Approximately 10.35\% of the scholars chose to lock their Twitter accounts, often signaling their departure from the platform by leaving a message in their profile directing followers to their Mastodon accounts. 

\begin{figure}[t]
    \centering
    \includegraphics[height=5.3cm]{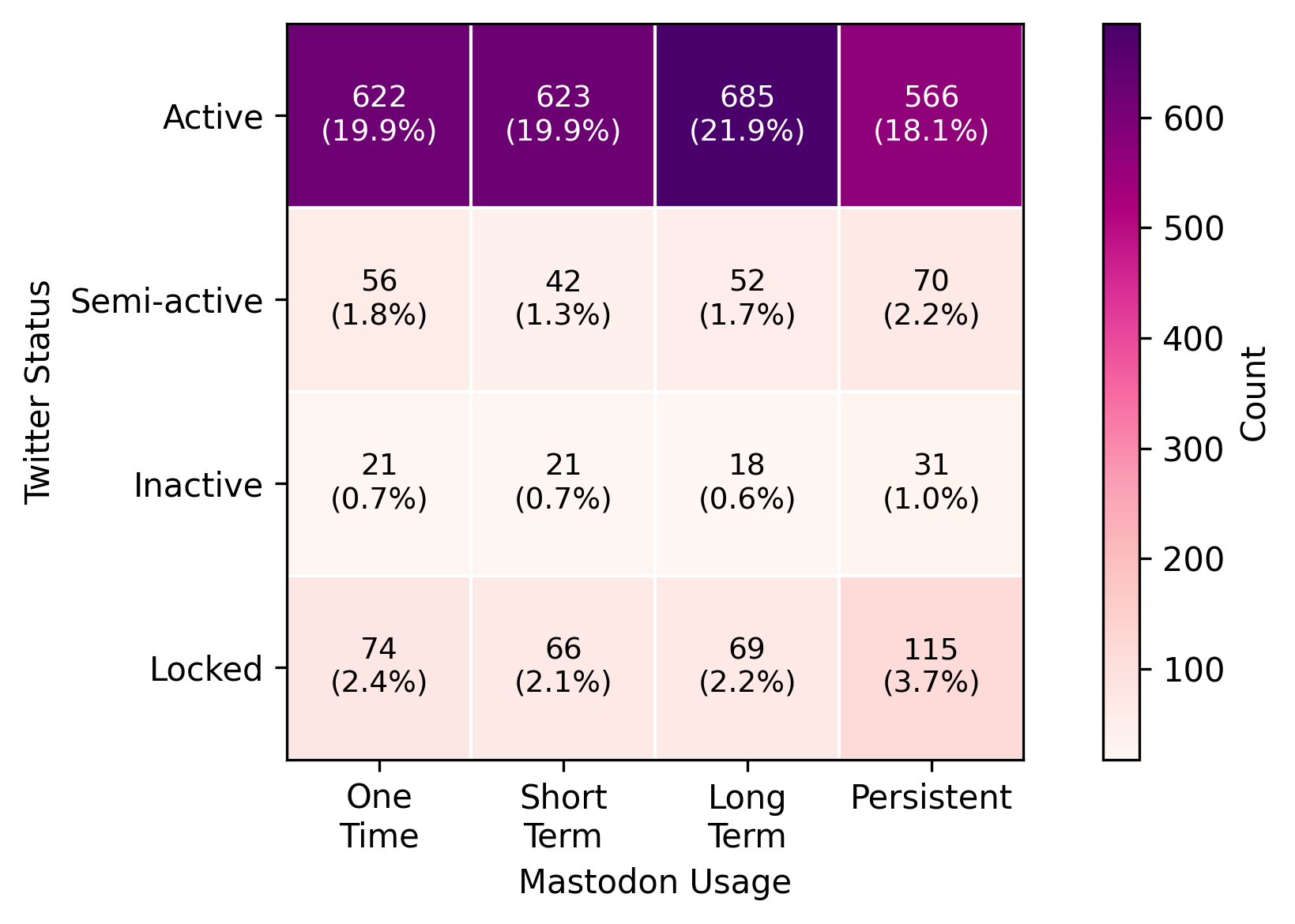}
    \caption{Cross-platform engagement mapping of Twitter and Mastodon; counts of scholars classified by Twitter status (rows) and Mastodon usage (columns).} 
    \label{fig:heatmap}
\end{figure}

We cross-mapped Twitter activity levels with Mastodon role categories, as shown in Figure \ref{fig:heatmap}. Our analysis reveals that 39.95\% of the sampled scholars are relatively active on both Mastodon (long-term or persistent users) and Twitter (active users). Among those who are inactive on Twitter or have locked their accounts, 56.14\% are long-term or persistent users of Mastodon. However, this group of dedicated Mastodon users represents only 7.44\% of the total scholars included in the cross-platform analysis. This suggests that while there is a core group of loyal users who have fully migrated to Mastodon, they make up a relatively small portion of the broader group of early academic adopters.

We also plotted the follower and following counts for the Mastodon and Twitter accounts of the same 3,131 scholars and the absolute differences in the metrics, as shown in Figure~\ref{fig:mapping}. The difference in connection scale is clear: while scholars who migrated to Mastodon made efforts to establish a new community, these networks remain far smaller than the connections many of them had built on Twitter over the years. Additionally, we plotted the Twitter account creation date against the total number of tweets, displayed in Figure~\ref{fig:Scatter}. The color of the points represents the total number of likes received by each account. A great proportion of scholars created their Twitter accounts between 2008 and 2012, indicating that the deep, long-standing networks, high levels of engagement, and rich posting histories developed on Twitter are not easily replicable or discarded. This entrenched presence highlights the difficulty of transitioning to a new platform like Mastodon without sacrificing years of accumulated influence and community ties.

In summary, we observe that the majority of scholars are either solely using Twitter or using both Twitter and Mastodon simultaneously. Comparing the follower and following counts across these platforms, Mastodon hosts a smaller community and Twitter continues to exert a strong pull on these early adopters.

\section{RQ2: Modeling Long-Term Retention}
In previous sections, we have identified multiple factors that could potentially influence persistence on Mastodon. To understand the effect of these factors on the persistence of Mastodon usage, we model them using a Cox proportional hazards model \cite{cox1972regression}. We selected this model to account for right censored data (users remaining active beyond the observation window) and to estimate the hazard ratios of covariates without assuming baseline hazard distribution.

\subsection{Mastodon Survival Analysis}
We used a Mastodon-centric model as our primary analysis and a cross-platform extension to understand the roles of Mastodon- and Twitter-derived features. The covariates used can be grouped as follows:

\vspace{0.1cm}
\noindent \emph{Mastodon activity metrics.} These represent users' platform activity level upon migration and social network metrics: the initial number of followers, following, and posts. 

\vspace{0.1cm}\noindent \emph{Federated interaction diversity.} To capture how broadly a user engages across the Fediverse beyond their home server, we compute a network diversity metric from interactions. For each user, we extract the set of outgoing interactions and map each interaction to a target server. We then compute Shannon entropy \cite{lin2002divergence} over the distribution of target servers, yielding higher values when a user’s outgoing interactions are spread across many servers rather than concentrated within a single server. Our primary model uses diversity measured over the first 30 days of observed activity and we report sensitivity checks using alternative definitions (14 days, 60 days, and the first 50 outgoing interactions) in  Figure~\ref{fig:model1_sensitivity} and \ref{fig:model2_sensitivity}.

\vspace{0.1cm}
\noindent \emph{Account and server-level features.} We capture users' profile settings, the network structure of their server, and the server type. Specifically, we consider whether the account is locked (binary), indicating if their posts are private and/or discoverable (binary), which indicates whether a user opted into being indexed to appear in search. At the server level, we identify the server type (binary) as either field-/topic-specific or general purpose 
by manually examining their public descriptions. A complete list of servers is provided in Appendix \ref{tab:topic_specific_servers}. At the server level, we define \emph{in-degree ratio} as the proportion of inbound connections relative to total connections and \emph{out-degree ratio} as the proportion of outbound connections relative to total connections. Additionally, we categorize users based on their academic discipline, creating a binary indicator for Multidisciplinary scholars in Table~\ref{tab:field_statistics} to test if these communities show a distinct retention patterns.

\vspace{0.1cm}
\noindent \emph{Cross-platform metrics.} These capture users' premigration history on Twitter. We consider the followers and following counts on Twitter, the total number of Tweets since account creation, and the age of their Twitter account in days. In both models, for each account, the number of days between our first data collection ({11-23-2022}) and the account's last recorded posting activity within the study window was used as survival time. We treat a user having experienced the event of becoming inactive if their last posting occurs before the end of our data collection ({10-27-2023}). The counts covariates are log transformed and z-score normalized to handle the covariate distribution and comparable coefficients. We used a penalization ($L_2=0.1$) to prevent overfitting. 

\subsection{Mastodon-centric Model} 
Our first model used the complete dataset of user accounts and features extracted exclusively from Mastodon data. We filtered out accounts on isolated servers and those with incomplete data, reducing our sample from 7,542 to a final account count of 7,357 users. The model achieved a strong predictive capability (concordance of 0.69) with several factors significantly predicting the risk of becoming inactive on Mastodon. The initial number of a user's followers ($HR=0.72,p<0.0005$) and number of posts ($HR=0.74,p<0.001$) were strongest protective factors; one standard deviation increase in either was associated with around 30\% lower likelihood of becoming inactive (refer to Appendix Figure \ref{fig:model1_sensitivity}). The initial number of accounts followed by a user did not have a significant effect ($HR=1.02,p=0.262$). 

For the server type and profile settings, being on a topic-specific server was a strong protective factor ($HR=0.84,p<0.0005$), aligning with our observations from the information security server that tailored communities are associated with higher user retention. Notably, users identified as Multidisciplinary scholars exhibited higher retention rates with a hazard ratio of $0.59$ $(p<0.001)$. Furthermore, having a discoverable profile was also significantly linked to lower attrition ($HR=0.93,p=0.009$), suggesting that being able to be found and connected is linked to persistent usage. In contrast, having the profile locked was not found to be a statistically significant predictor. The out-degree ratio was a risk factor ($HR=1.80,p<0.001$) suggesting servers that are heavily outward looking are associated with lower user persistence. However, the proportion of inbound connections was not a significant predictor. 

Federated interaction diversity is a robust protective predictor. In the 30-day diversity model, network diversity is associated with lower attrition ($HR=0.84,p<0.001$), indicating that a one standard deviation increase in early cross-server interaction diversity corresponds to an approximately 16\% lower hazard of becoming inactive. Sensitivity checks using alternative windows (14 days, 60 days, and first 50 outgoing interactions) yield consistent results (Figures \ref{fig:model1_sensitivity}), with effect sizes ranging from $HR=0.89$ (14 days) to $HR=0.75$ (first 50 outgoing interactions).

\subsection{Cross-platform Model} 
Our second model was fitted using the subset of 3,131 users for whom we had complete cross-platform features from both Mastodon and Twitter. This allowed us to understand how a user's Twitter activity related to their retention on Mastodon. The cross-platform model achieved concordance of 0.69 indicating that the addition of Twitter features does not add much overall predictive power. 
The addition of Twitter covariates provides a granular view of retention correlates. Notably, the association with interaction network diversity remains robust ($HR=0.86,p<0.001$), indicating that cross-server engagement predicts retention independent of a user's prior status on Twitter.

\begin{figure}[ht]
    \centering
    \includegraphics[width=\linewidth]{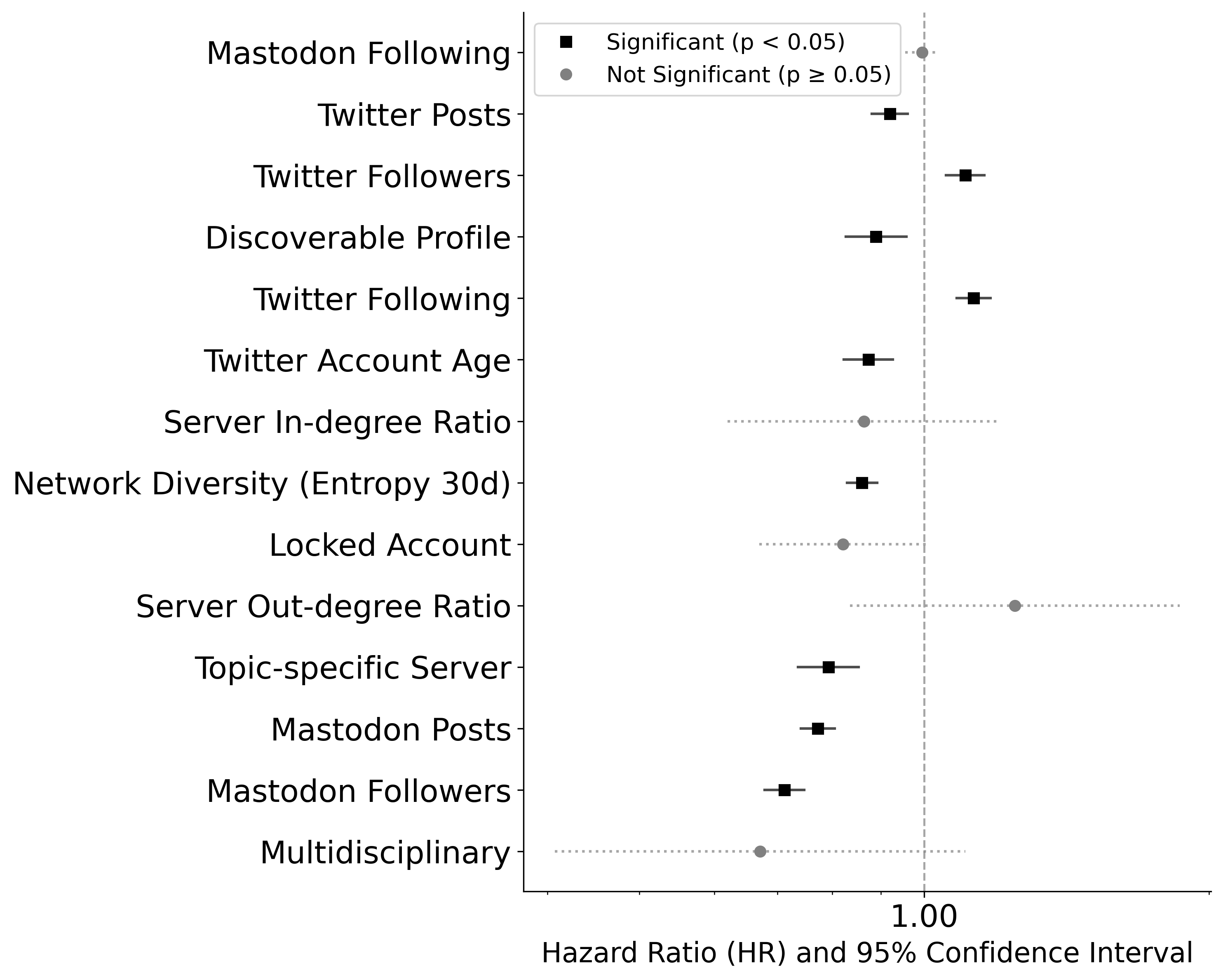}
    \caption{Plot showing hazard ratio and 95\% CI for each covariate in the cross-platform model. Vertical dashed line at HR of 1.0 represents no effect.}
    \label{fig:model2}
\end{figure}

Among covariates used, the association of metrics extracted from Mastodon remained largely consistent with the first model (see Figure \ref{fig:model2}). Initial followers on the platform ($HR=0.72,p<0.0005$) and number of posts ($HR=0.77,p<0.001$) were still strong factors linked to retention. Similarly, server type ($HR=0.79,p<0.001$) and discoverability ($HR=0.86,p<0.001$) remained significant and positively associated with persistence. However, unlike the Mastodon-centric model, server structural metrics and Multidisciplinary scholar indicator were no longer significant predictors of attrition in the cross-platform model.

Results of modeling covariates from Twitter revealed nuanced effects of some of the cross-platform user characteristics on persistence. A large number of Twitter following ($HR=1.13,p<0.0005$) and followers ($HR=1.11,p<0.0005$) were significant risk factors for becoming inactive on Mastodon. In contrast, a higher number of Twitter posts ($HR=0.92,p<0.001$) and older Twitter account age ($HR=0.88,p<0.0005$) were linked to higher retention. 

The results from the cross-platform analysis shows a clear relationship with Twitter history. A large Twitter network is associated with higher attrition, while a long history of Twitter activity is linked to persistence showing divergent associations for these factors. Users with a well-established presence and high posting volume on Twitter were more committed to migration, but the size of their network may have been difficult to replicate on Mastodon, pulling them back to Twitter. This suggests that a successful migration of a community from a well-established platform like Twitter requires not only individual commitment but also a critical mass of connections that can migrate to the new platform.

\section{Contextual Factors and Platform Environment}

The models identify measurable predictors of sustained use, yet migration and retention dynamics also unfold within a broader contextual environment that is not fully observable from activity traces alone. In this section, we examine three such contextual factors: the emergence of competing alternative platforms, public discourse surrounding the academic migration, and Mastodon’s platform affordances. These factors help contextualize the conditions under which early adopters navigated migration and sustained participation.

\subsection{Alternatives to Mastodon}
In addition to Mastodon, other centralized alternatives to Twitter have emerged in recent years, and many academics explored these venues. Unlike Mastodon, these platforms provided familiar interfaces and ease of transitioning. Those present on these alternative platforms often added their account information in their Twitter account descriptions to make it easier for others to discover and connect with them on these new platforms. To understand the extent to which academics were exploring the alternatives, we analyzed their usernames and user descriptions for mentions of these platforms. Specifically, we focused on mentions of Bluesky\footnote{\url{https://bsky.app/}} and Threads\footnote{\url{https://www.threads.net/}} using keyword matching as well as user name pattern matching. For Mastodon and Bluesky, we also included scholars who use decentralized platform-style usernames as their Twitter usernames (e.g., $<$username$>$@$<$servername$>$).

\begin{figure}[ht]
    \centering
    \includegraphics[width=0.8\linewidth]{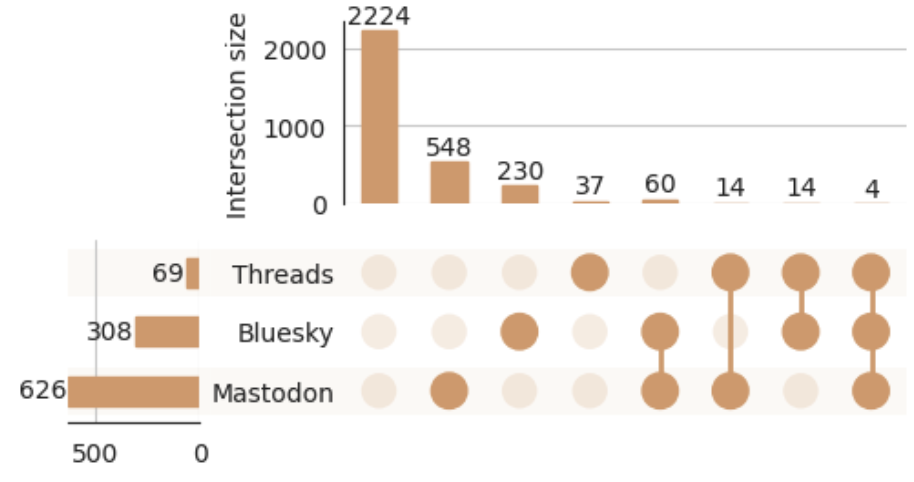}
    \caption{UpSet plot summarizing explicit cross-platform self-identification; intersection bars indicate counts of scholars mentioning Mastodon, Bluesky, and Threads (alone or jointly) in their profiles.} 
    \label{fig:platforms}
\end{figure}

Figure~\ref{fig:platforms} shows an UpSet plot illustrating the number of users who explicitly mention each platform or a combination of platforms. In addition to 626 scholars mentioning Mastodon, 308 and 69 users also reference BlueSky and Threads, respectively, with some scholars providing account information for multiple platforms. This indicates that while these scholars have accounts on Mastodon, only about 20\% actively direct their Twitter connections to it. The trend highlights that Mastodon is not the sole option for scholars migrating away from Twitter and faces significant competition from other emerging platforms. This competition further complicates Mastodon's ability to establish itself as the primary platform for academic discourse.

\subsection{Public Discourse on Academic Migration}
\begin{figure}[ht]
    \centering
    \includegraphics[width=\linewidth]{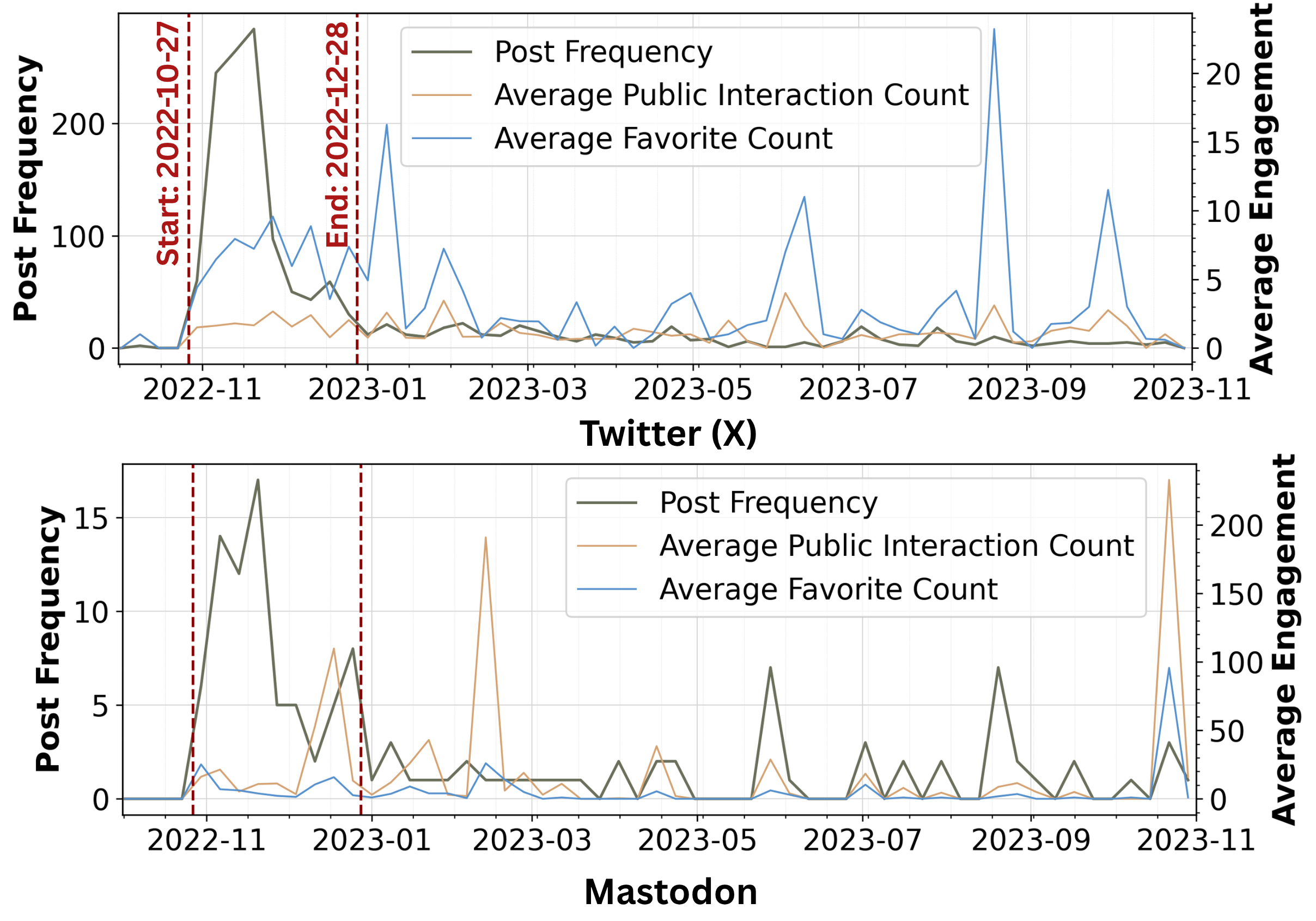}
    \caption{Dynamic frequency of Twitter and Mastodon conversations related to academics migration and corresponding engagement metrics (public interaction and favorite counts), aggregated by week.}
    
    \label{fig:dyn_Twitter}
\end{figure}

To examine public discourse surrounding early academic migration, we analyze public posts from Twitter and Mastodon that explicitly reference academic migration, collected using a predefined keyword-based retrieval strategy (see Section Data). We plot the daily frequency of these posts and engagement counts over time aggregated by week in Figure~\ref{fig:dyn_Twitter}. We observe similar patterns on both platforms: discussions about the migration in the academic context experienced a noticeable spike during the first two months of the movement on both platforms. After the migration period, conversations became less frequent on both platforms. However, they received more attention (favorites) on Twitter, indicating that the audience and connections of academics concerned about Mastodon largely remained active on Twitter.

Qualitative inspection of highly engaged posts during peak periods reveals recurring themes related to uncertainty about platform stability and long-term viability. For example, Mastodon struggled to accommodate the abrupt influx of users. On the peak day of discussion, November 18, 2022:
\begin{quote}

    \textit{``I have had a Mastodon presence for several months but haven't used it. I still have a vain hope Twitter will sort itself out, but if not, I am $<$scholar's Mastodon account$>$."} --Twitter Post

\end{quote}

\noindent However, this initial surge in discussion was short-lived. The frequency of related tweets sharply declined. 
Users expressed frustration over the platform's lack of centralized theme and purpose, and its fragmented community structure.

\begin{quote}

    \textit{``My humanities folks have largely migrated to mastodon, but my science and public health feeds are still here on Twitter. More’s the pity." } --Twitter Post
\end{quote}

Despite these challenges, public efforts to encourage a shift toward Mastodon continued throughout 2023. For example, one scholar highlighted its advantages in a June 2023 post on Twitter:
\begin{quote}
    \textit{``Reasons to join us in the fediverse. 500 character limit. free edit button. Lots more \#openaccess resources for \#teachers, researchers, science communicators, lifelong learners. \#STEM \#education \#openscience \#K12 \#HigherEd"} --Twitter Post
\end{quote}

Momentum was also visible within Mastodon itself. A widely circulated post from October 2023, which received 613 reblogs, signaled a clear departure from Twitter:

\begin{quote}
    \textit{Farewell, [bird emoji]! After 14 years on Twitter/X, we feel it's time to part ways. Recent developments there do not align with our core values. We stand for trustworthy information provision and are committed to open science. Follow us here for open science-related updates and news on our services and support for the academic community.[heart emoji] \#OpenScience \#AcademicLibrary \#GoodbyeTwitter \#ResearchSupport \#EducationSupport} --Masotdon Post 
\end{quote}

\noindent Amongst the discussions on Twitter, we again observe concurrent mentions of other platforms such as Bluesky and Threads. These have diluted the migration as users explore various platforms for different purposes. For instance,
\begin{quote}

    \textit{``Hey if anyone's still here on science Twitter and wondering where a lot of people have gone, they're trying out bluesky and liking it. This time it's real (not like mastodon). \#academic"} --Twitter Post

\end{quote}
Taken together, these posts illustrate a mix of optimism about Mastodon’s potential and recurring practical concerns users have faced during the migration.

\subsection{Mastodon's Technical Affordances}

Beyond the individual- and community-level factors identified above, structural characteristics of Mastodon, such as its moderation policies and discoverability mechanisms, can be considered additional factors contributing to the limited success of migration.
In line with the network structure, moderation policy is also decentralized, as described\footnote{https://docs.joinmastodon.org/admin/moderation/}:  
\begin{quote}
    \textit{Moderation in Mastodon is always applied locally, i.e. as seen from the particular server. An admin or moderator on one server cannot affect a user on another server, they can only affect the local copy on their own server.}
\end{quote}

This decentralized moderation stands in sharp contrast to Twitter’s centralized enforcement system. While local control empowers individual communities to define and enforce their own standards, it also creates inconsistencies across the network. For users accustomed to a uniform, centralized policy environment such as Twitter’s, these variations may generate uncertainty or limit trust in the platform.

Discoverability presents another technical challenge as shown in the models. Unlike Twitter, which supports platform-wide search and algorithmic amplification of content, Mastodon limits search to hashtags and profile information\footnote{https://docs.joinmastodon.org/user/network/\#search}. As a result, newcomers may struggle to discover communities, topics, or influential users, creating barriers to community formation and sustained engagement.

Taken together, these limiting affordances could help explain why Mastodon, despite its alignment with values of autonomy and community governance, has struggled to attract and retain large numbers of migrating users from Twitter. 
Importantly, these technical affordances are not experienced uniformly across the Fediverse. Because each Mastodon server operates with its own administrative history, governance structure, and moderation capacity, users’ experiences may vary substantially depending on the server they join. Servers established prior to the migration period may benefit from more mature moderation practices or greater administrative experience, which could influence user satisfaction and retention.

\section{Discussion and Conclusions}

In this work, we explored the migratory dynamics of Academic Twitter communities to Mastodon. Through a series of analyses, we highlight that building a community on a new platform requires more than a coordinated effort and availability of alternatives. 
Results from longitudinal analysis showed that the initial surge in sign-ups did not translate into sustained long-term user engagement. This pattern is consistent with prior studies on Twitter to Mastodon migration in general user groups showing early migration waves often do not translate to stable long-term activity with many users maintaining cross-platform presence and continue using Twitter as a primary platform for engagement rather than fully switching \cite{jeong2023exploring, he2023flocking} and with broader findings that the transition away from Twitter remains incomplete for general audiences \cite{pena2025straddling}. Most recent user activity on Twitter revealed that many academics in our dataset returned to using Twitter. Further analysis suggested that long-standing histories and strong communities on Twitter, in some cases spanning over a decade, proved too significant to overcome. While prior studies observed this general return to Twitter \cite{jeong2023exploring,he2023flocking}, our cross-platform analysis with Twitter features suggests that, for established scholars, the difficulty of replicating existing networks is correlated with attrition. Without a complete migration of user history and established community, professionals and other specialized groups may naturally gravitate back to familiar platforms where their communities are already robust. 

Likewise, several factors were linked to the failure of Mastodon to retain users. First, its decentralized structure creates a steeper learning curve compared to Twitter. Competition from more user-friendly platforms like Bluesky and Threads perhaps drew users away from Mastodon. Beyond direct Twitter substitutes, professional platforms such as LinkedIn continue to serve distinct functions centered around academic communication \cite{utz2019relationship}, which may further diffuse scholars' attention across platforms rather than concentrating migration on a single alternative. Our analysis adds to the understanding of challenges in migration of professional communities by identifying interaction diversity as a robust predictor of retention. Additionally, those who continued to post consistently on Mastodon did not enjoy the same level of engagement as on Twitter. That being said, select groups of academics stayed committed to Mastodon, making this migration less of a straightforward failure and more of a complex transition. Prior work suggests that social influence and a shared sense of identity or belonging can shape collective migration behavior \cite{cava2023drivers}. In our data, higher persistence among scholars on field or topic-specific servers is consistent with the possibility of community-specific servers supporting sustained engagement. This commitment is likely reinforced by self selection. For example, the most motivated early adapters may have actively sought out community-specific servers.

Future efforts that support migrations to decentralized platforms must address these core issues, foster community building by reducing friction, and consider network effects on user engagement. While the idea of decentralization of social media platforms holds significant promise, practical challenges persist that require more than just technical solutions. 
The Academic Twitter community's large-scale attempt to migrate from Twitter to Mastodon is significant--a strong professional network with a shared purpose  positioned it as a hopeful candidate to succeed with such migration. Yet, as encouraging as these efforts and their partial successes are, evidence suggests that the process of transition is extremely challenging.

\section{Limitations}

One limitation of this study is inconsistency in the timing of data collection across platforms. This discrepancy could potentially introduce biases in the analysis, as activity levels of scholars might vary depending on when the data was captured. Additionally, incorrect or incomplete social media handles provided by the scholars led to gaps in data that we could not manually resolve due to the scale of the dataset. While this subset provides substantial coverage of early adopters and enables examination of cross-platform engagement behaviors within this group, the analysis is descriptive in nature and not intended to support population-level claims. Moreover, changes to platform access (e.g., Twitter API rate limitations) restricted our ability to gather complete data throughout the study. Similarly, full text keyword search is unavailable on many Mastodon instances unless it is enabled by the server administrator, so our reliance on hashtag-based search, which works without special server configuration, may have resulted in a smaller set of retrievable posts. In addition, the keywords used to collect public discourse may not capture all forms of academic self-description, which introduces the possibility of missing some relevant posts. However, we believe our findings are nevertheless substantial and provide valuable insights for cross-platform analyses. 
Notably, collected data reflects early groups of academic migrants, which are likely more indicative of collective migration efforts. However, we acknowledge the possibility of selection bias, as these patterns may not represent later migrants. This focus on early adopters implies that our findings primarily capture the initial, coordinated dynamics of migration rather than the full range of user experiences. Furthermore, our analysis treats a user's server as static which does not account for potential between-server migration within the Fediverse after the server choice at account creation.
Finally, there may be case-specific factors at both the individual and community levels, including differences in local governance, historical development, or disciplinary norms, that cannot be captured holistically through platform traces alone. While such factors fall outside the scope of this study, we believe the observed patterns nevertheless provide a useful empirical account of early academic migration dynamics.

\section{Ethical Considerations}

All account IDs in the original curated lists were voluntarily added by the users. Our data collection was restricted to publicly available interaction and profile metadata on Mastodon and Twitter at the time of collection. No private posts, direct messages, or other non-public information were accessed or retained. All identifiable information has been removed from the paper and associated research artifacts to further protect user privacy. Given the relatively narrow nature of the data collected in this study, we believe the risk of privacy violation and misuse to be minimal.


%
%
%
%

\bibliography{main.bib}

\subsection{Paper Checklist}

\begin{enumerate}
\item General
\begin{enumerate}
    \item  Would answering this research question advance science without violating social contracts, such as violating privacy norms, perpetuating unfair profiling, exacerbating the socio-economic divide, or implying disrespect to societies or cultures?
    Yes, see Section Ethical Consideration. 
  \item Do your main claims in the abstract and introduction accurately reflect the paper's contributions and scope?
    Yes.
   \item Do you clarify how the proposed methodological approach is appropriate for the claims made? 
    Yes, we detail the approaches used in Research Methods section.
   \item Do you clarify what are possible artifacts in the data used, given population-specific distributions?
    NA. In this work, we are focusing on a group of online users who participated within a coordinated migration. And we have no intention to generalize the claims to the overall population.
  \item Did you describe the limitations of your work?
    Yes, see Limitation section.
  \item Did you discuss any potential negative societal impacts of your work?
    Yes, please refer to the Ethical Considerations section of the article.
      \item Did you discuss any potential misuse of your work?
    Yes, please refer to the Ethical Considerations section of the article.
    \item Did you describe steps taken to prevent or mitigate potential negative outcomes of the research, such as data and model documentation, data anonymization, responsible release, access control, and the reproducibility of findings?
    Yes, Please refer to the Ethical Considerations section of the article.
  \item Have you read the ethics review guidelines and ensured that your paper conforms to them?
    Yes.
\end{enumerate}

\item Additionally, if your study involves hypotheses testing...
\begin{enumerate}
  \item Did you clearly state the assumptions underlying all theoretical results?
    NA.
  \item Have you provided justifications for all theoretical results?
    NA.
  \item Did you discuss competing hypotheses or theories that might challenge or complement your theoretical results?
    NA.
  \item Have you considered alternative mechanisms or explanations that might account for the same outcomes observed in your study?
    NA.
  \item Did you address potential biases or limitations in your theoretical framework?
    NA.
  \item Have you related your theoretical results to the existing literature in social science?
    NA.
  \item Did you discuss the implications of your theoretical results for policy, practice, or further research in the social science domain?
   NA.
\end{enumerate}

\item Additionally, if you are including theoretical proofs...
\begin{enumerate}
  \item Did you state the full set of assumptions of all theoretical results?
    NA.
	\item Did you include complete proofs of all theoretical results?
    NA.
\end{enumerate}

\item Additionally, if you ran machine learning experiments...
\begin{enumerate}
  \item Did you include the code, data, and instructions needed to reproduce the main experimental results (either in the supplemental material or as a URL)?
    NA.
  \item Did you specify all the training details (e.g., data splits, hyperparameters, how they were chosen)?
    NA.
     \item Did you report error bars (e.g., with respect to the random seed after running experiments multiple times)?
    NA.
	\item Did you include the total amount of compute and the type of resources used (e.g., type of GPUs, internal cluster, or cloud provider)?
    NA.
     \item Do you justify how the proposed evaluation is sufficient and appropriate to the claims made? 
    NA.
     \item Do you discuss what is ``the cost`` of misclassification and fault (in)tolerance?
    NA.
  
\end{enumerate}

\item Additionally, if you are using existing assets (e.g., code, data, models) or curating/releasing new assets, \textbf{without compromising anonymity}...
\begin{enumerate}
  \item If your work uses existing assets, did you cite the creators?
    Yes.
  \item Did you mention the license of the assets?
    NA.
  \item Did you include any new assets in the supplemental material or as a URL?
    NA.
  \item Did you discuss whether and how consent was obtained from people whose data you're using/curating?
    No, in this study we are not releasing any new data with personally identifiable information.
  \item Did you discuss whether the data you are using/curating contains personally identifiable information or offensive content?
    NA.
\item If you are curating or releasing new datasets, did you discuss how you intend to make your datasets FAIR?
NA.
\item If you are curating or releasing new datasets, did you create a Datasheet for the Dataset? 
NA.
\end{enumerate}

\item Additionally, if you used crowdsourcing or conducted research with human subjects, \textbf{without compromising anonymity}...
\begin{enumerate}
  \item Did you include the full text of instructions given to participants and screenshots?
    NA.
  \item Did you describe any potential participant risks, with mentions of Institutional Review Board (IRB) approvals?
    NA.
  \item Did you include the estimated hourly wage paid to participants and the total amount spent on participant compensation?
    NA.
   \item Did you discuss how data is stored, shared, and deidentified?
   NA.
\end{enumerate}

\end{enumerate}

\newpage

\onecolumn
\section{Appendix}
\renewcommand{\thetable}{A\arabic{table}}
\renewcommand{\thefigure}{A\arabic{figure}}
\setcounter{table}{0} 
\setcounter{figure}{0}
\scriptsize

\scriptsize
\setlength\LTleft{\fill}
\setlength\LTright{\fill}

\begin{table*}[h]
\centering
\scriptsize
\begin{tabular}{l|ll|l}
\toprule
Field & Discipline & Count & Total \\
\midrule
\multirow{13}{*}{Arts and Humanities}&Archaeology&126&\multirow{13}{*}{1522}\\
&Art and Design&8&\\
&Biblical Studies&4&\\
&Digital Humanities&208&\\
&Glams (Galleries, Libraries, Archives, Museums, Special Collections)&109&\\
&Literature Studies&142&\\
&Philosophy&220&\\
&Religion&56&\\
&Theology&27&\\
&History&554&\\
&History and philosophy of science &11&\\
&BookHistory&36&\\
&Medieval studies&21&\\ 
\midrule
Business&Marketing&4&4\\
\midrule
Education&Education&244&244\\

\midrule
Law&Law&587&587\\
\midrule

\multirow{4}{*}{Life Sciences}&Bioinformatics&13&\multirow{4}{*}{84}\\
&Biophysics&8&\\
&Palaeogenomics&6&\\
&Plant Science&57&\\
\midrule

\multirow{2}{*}{Medicine and Health Sciences}&Neuroscience&264&\multirow{2}{*}{287}\\
&Medical AI&23&\\

\midrule
\multirow{9}{*}{Physical Sciences and Mathematics}&Chemistry&240&\multirow{9}{*}{851}\\
&Cheminformatics and Computational Chemistry&11&\\
&Geology&145&\\
&Meteorology&93&\\
&Nuclear fusion&1&\\
&Astrophysics&53&\\
&Atmospheric and air quality&41&\\
&Dendrochronology&3&\\
&Information Security&264&\\
\midrule

\multirow{17}{*}{Social and Behavioral Sciences}&Anthropology&50&\multirow{17}{*}{2918}\\
&Communication and Media Studies&265&\\
&Criminology&41&\\
&Genealogy&168&\\
&Geography&282&\\
&Linguistics&108&\\
&Political Science&364&\\
&Sociology&228&\\
&Asian Studies&26&\\
&Iran and middle eastern studies &5&\\
&Psychology&321&\\
&Public Policy&30&\\
&Technology Policy&152&\\
&Science and technology studies &41&\\
&Semantic Web&25&\\
&Qualitative Research&35&\\
&Journalism&777&\\

\midrule
\multirow{2}{*}{Multi-disciplinary} & Multi-disciplinary & 961 & \multirow{2}{*}{1040} \\
&Open Science&79&\\
\bottomrule
\end{tabular}

\caption{Number of scholars by disciplines within major academic fields.}
\label{tab:field_statistics}
\end{table*}
\begin{figure*}[h!]
    \centering
    \includegraphics[width=\linewidth]{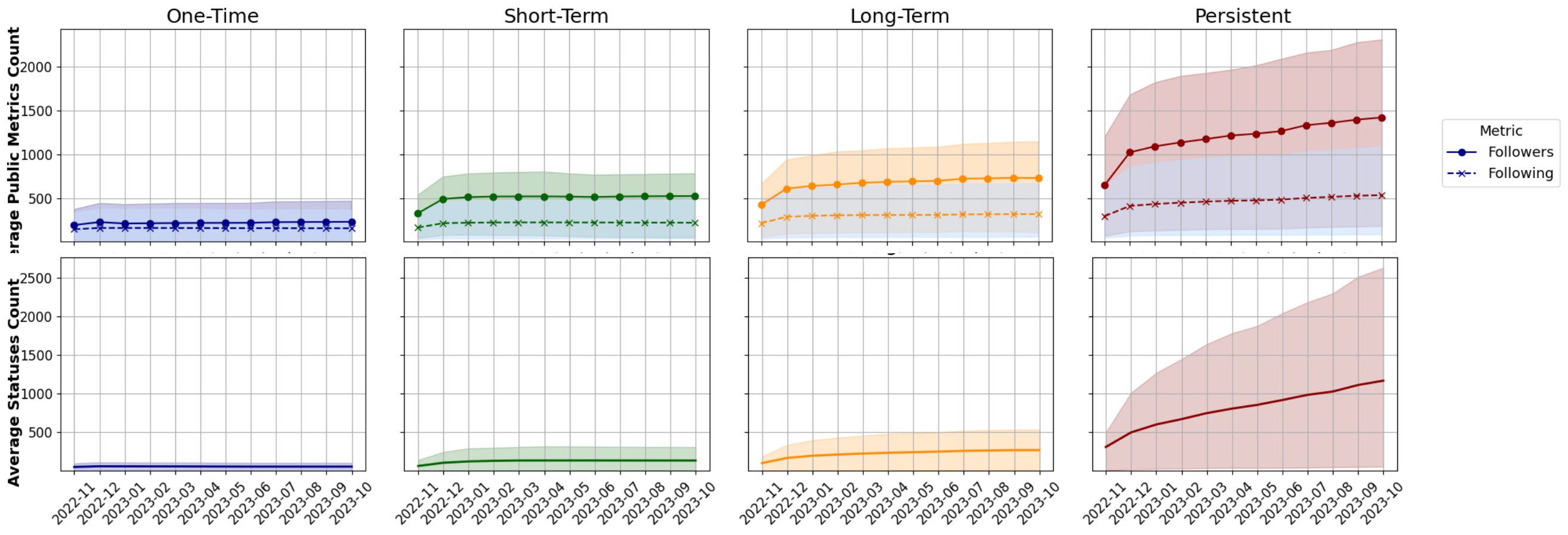}
    \caption{Monthly average total public metrics and statuses count for active users with  10–90\% Range.} 
    \label{fig:metrics}
\end{figure*}

\begin{figure*}[h!]
    \centering
    \includegraphics[width=0.8\linewidth]{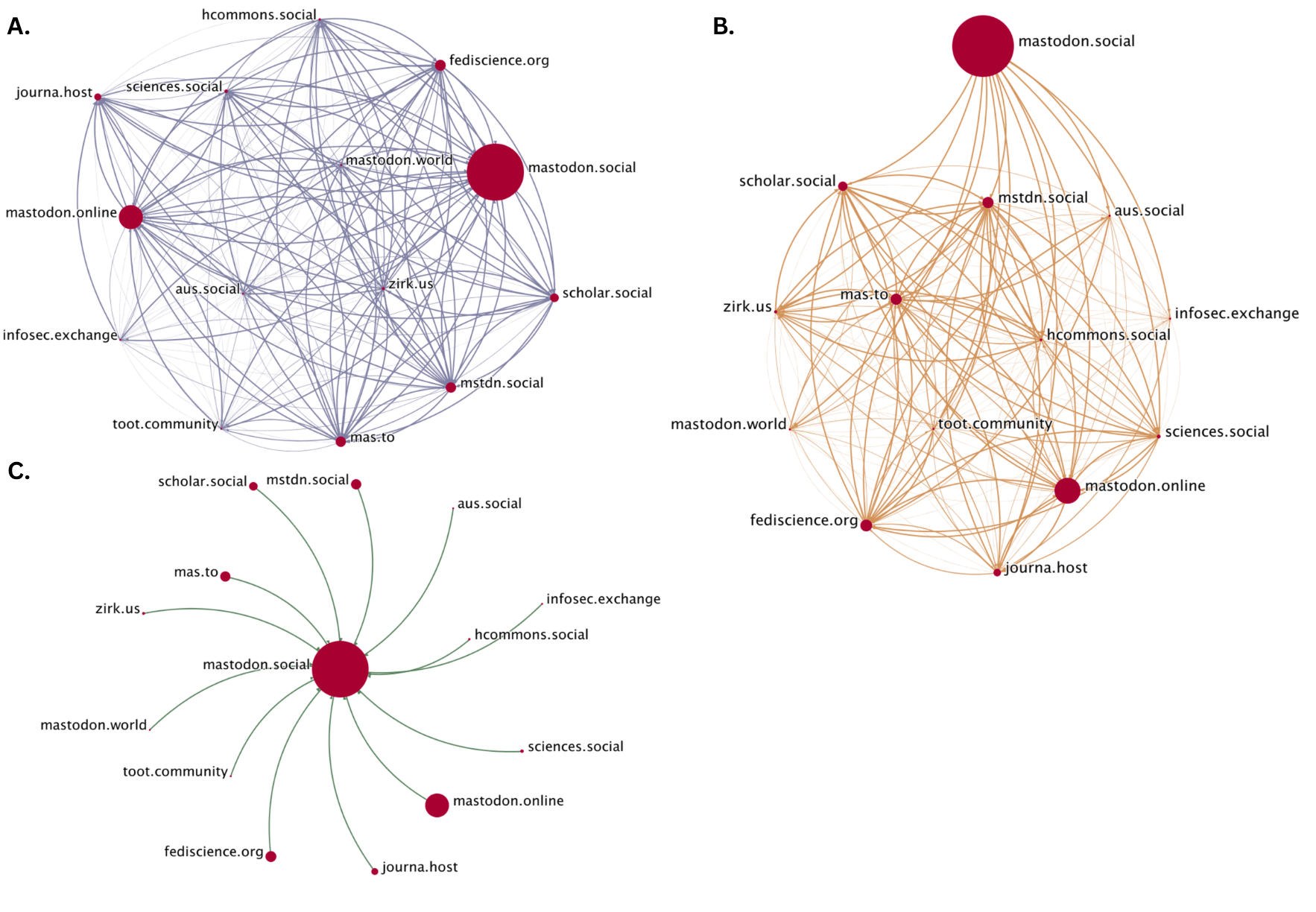}
    \caption{Internal interaction networks among servers with more than 100 scholars. A: boost; B: mention; C: favorite. }
    \label{fig:rest}
\end{figure*}

\begin{figure*}[h!]
    \centering    \includegraphics[width=0.5\linewidth]{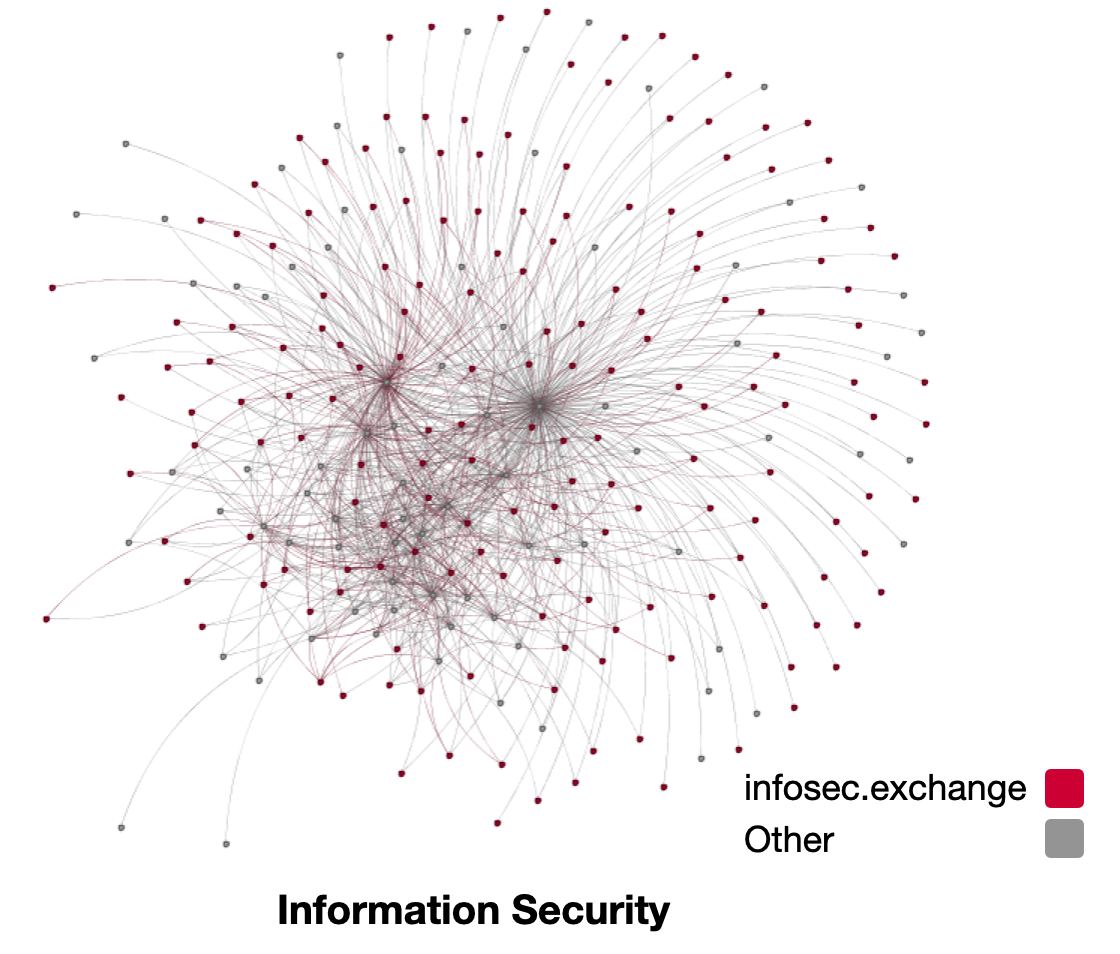}
    \caption{internal follower-followee network for Information Security community (removed isolates for better visualization).}
    \label{fig:infosec}
\end{figure*}
\begin{figure*}[h!]
    \centering
    \includegraphics[height=7.45cm]{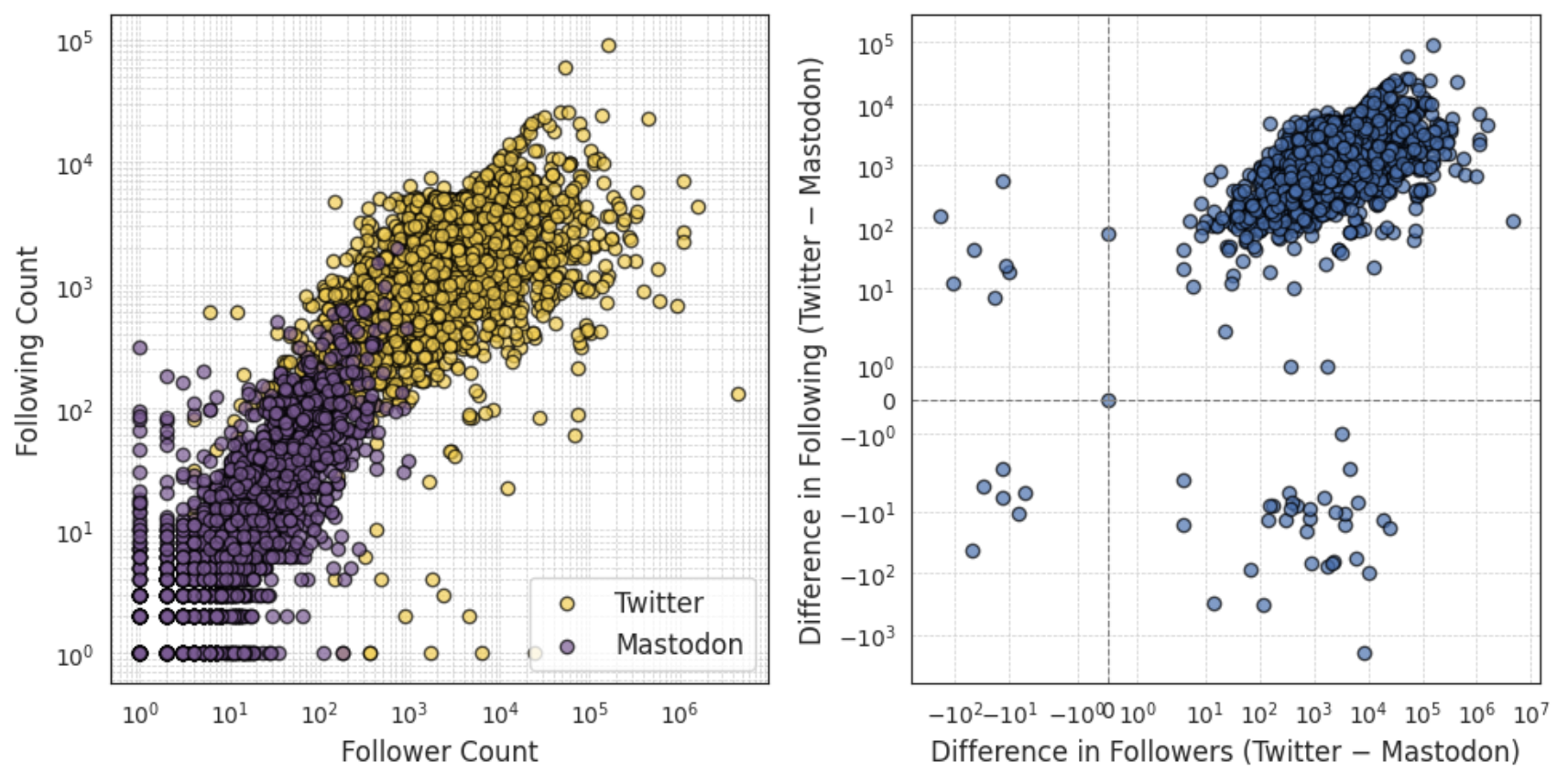}
    \caption{Connection metrics mapping and differences in metrics between mastodon and Twitter. }
    \label{fig:mapping}
\end{figure*}

\begin{figure*}[h!]
    \centering
    \includegraphics[width=\linewidth]{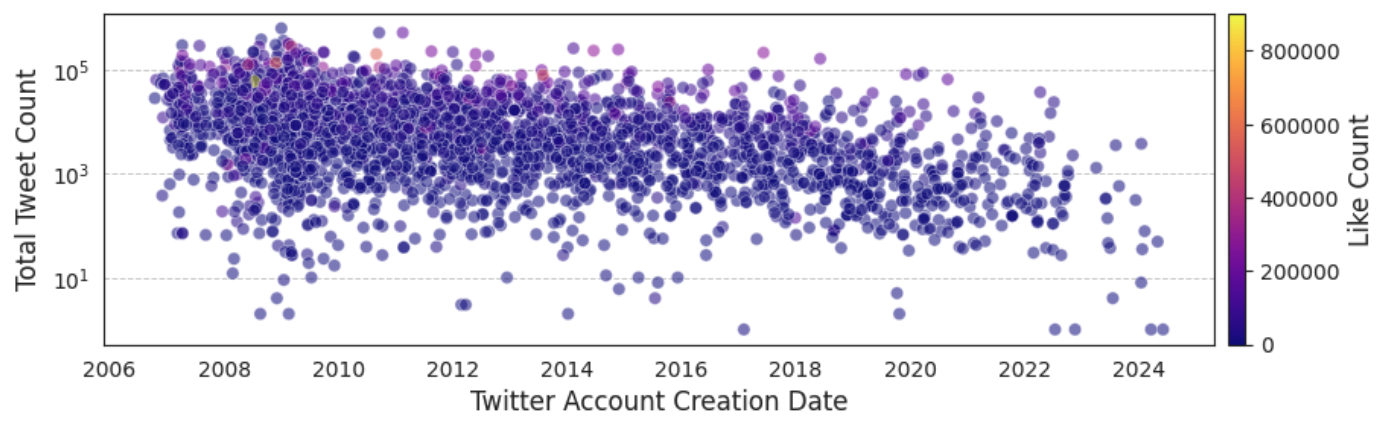}
    \caption{Scatter Plot of Scholar's Twitter Account Creation Date VS. Engagement Metrics. }
    \label{fig:Scatter}
\end{figure*}

\begin{table*}[h]
\tiny
\centering
\resizebox{\columnwidth}{!}{
\fontsize{7.5}{11}\selectfont{
\begin{tabular}{l|cccc|cccc}
\toprule
\textbf{Field} & \textbf{One Time(\%)} & \textbf{Short Term(\%)} & \textbf{Long Term(\%)} & \textbf{Persistent(\%)} & \textbf{Total} & \textbf{Percent} & \textbf{In-Degree} & \textbf{Out-Degree}\\
\midrule
Social \& Behavioral Sciences &  25.0&  25.8&  24.1&  25.1& 2918 & 38.72 & 98881 & 91958\\
Arts \& Humanities&  29.0& 21.4&  26.0&  23.6& 1522 & 20.19 & 38692 & 39986\\
Physical Sciences \& Mathematics &  22.1&  23.4&  27.6& 26.9& 851 & 11.29 & 11453 & 11980\\
Life Sciences &  20.2&  22.6&  35.7&  21.4& 84 & 1.11 & 338 & 648\\
Education &  32.8& 25.0&  20.5&  21.7 & 244 & 3.24 & 2249 & 4171\\
Law &  29.6& 28.3& 25.4&  16.7 & 587 & 7.79 & 21664 & 27450\\
Medicine \& Health Sciences &  24.7&  31.0&  26.8 & 17.4& 287 & 3.81 & 4292 & 4923\\
Multidisciplinary & 18.6& 25.1&  27.1& 29.2& 1040 & 13.80 & 56371 & 52818\\
Business & 25.0&50.0&0.0& 25.0& 4 & 0.05 & 60 & 66 \\
\bottomrule
\end{tabular}}}
\caption{User activity patterns by academic field (one-time, short-term, long-term, persistent) and field-level statistics (total users, percentage, and network centrality indicators).}
\label{tab:percentage}
\end{table*}

\begin{table*}[h]
\centering
\small
\resizebox{\columnwidth}{!}{
\begin{tabular}{lcccc|p{6cm}|c}
\toprule
Server & Count & Percentage& In-Degree & Out-Degree & Description &Scope \\
\midrule
mastodon.social & 1606 &21.31\%& 147295 & 128525 &The original server operated by the Mastodon GmbH non-profit.&Universal \\
mstdn.social & 538 &7.14\%& 9456 & 11947 &A general-purpose Mastodon server with a 500 character limit. All languages are welcome.&Universal \\
mastodon.online & 429 &5.69\%& 6600 & 7172 &A newer server operated by the Mastodon GmbH non-profit.&Universal \\
fediscience.org & 415 &5.51\%& 8220 & 9628 &Fediscience is the social network for scientists. &Universal\\
mas.to & 381 &5.06\%& 5591 & 6534 &Hello! mas.to is a fast, up-to-date and fun Mastodon server.&Universal\\
sciences.social & 218 &2.89\%& 5535 & 7460 &Non-profit, ad-free social media for social scientists. Join thousands of social scientists here and across the fediverse.  &Field-level\\
infosec.exchange & 200 &2.65\%& 1001 & 1362 &A Mastodon instance for info/cyber security-minded people. &Discipline-level\\
scholar.social & 193 &2.56\%& 2842 & 3899&Microblogging for researchers, grad students, librarians, archivists, undergrads, high schoolers, educators, research assistants, profs—anyone involved in learning who engages with others respectfully. &Universal\\
journa.host & 165 &2.19\%& 3086 & 2717&The server for working journalists and news outlets on Mastodon. Home to active \& retired journalists, media scholars, and a variety of news and journalism adjacent professionals. \#Newstodon &Discipline-level\\
zirk.us & 153 &2.03\%& 2276 & 3087&Literature, philosophy, film, music, culture, politics, history, architecture: join the circus of the arts and humanities! For readers, writers, academics or anyone wanting to follow the conversation. &Field-level \\
mastodon.world & 144 &1.91\%& 1655 & 2213 &Generic Mastodon server for anyone to use. &Universal\\
toot.community & 107 &1.42\%& 805 & 1202&A worldwide Mastodon instance from The Netherlands. Run by digital enthusiasts, inviting everyone, everywhere to join us in the \#fediverse.&Universal  \\
hcommons.social & 121 &1.61\%& 2110 & 2363&hcommons.social is a microblogging network supporting scholars and practitioners across the humanities and around the world.&Universal \\
aus.social & 101 &1.34\%& 1156 & 1255&Welcome to thundertoot! A Mastodon Instance for the People. &Universal \\
\bottomrule
\end{tabular}
}
\caption{Centrality metrics for Mastodon servers with more than 100 scholars, ordered by count.}
\label{tab:server_centrality_metrics}
\end{table*}

\begin{table*}[h]
\centering
\small
\begin{tabular}{|p{17cm}|}
\toprule
\textbf{Field/Topic specific servers}\\
\midrule
infosec.exchange, journa.host, scholar.social, fediscience.org, sciences.social, zirk.us,
hcommons.social, astrodon.social, drosophila.social, fosstodon.org, qoto.org, masto.ai,
genomic.social, hci.social, mathstodon.xyz, med-mastodon.com, ecoevo.social, bayes.club,
dair-community.social, mapstodon.space, sigmoid.social, universeodon.com, sciencemastodon.com,
neuroscience-mastodon.com, datasci.social, meteo.social, mstdn.science, ioc.exchange, wxcloud.social,
skew-t.com, eupolicy.social, law.builders, mastodon.lawprofs.org, esq.social, legal.social,
publiclaw.space, openbiblio.social, ausglam.space, lingo.lol, archaeo.social, medievalist.masto.host,
econtwitter.net, digipres.club, glammr.us, c18.masto.host, religion.masto.host, fedihum.org,
bildung.social, book\_historia, coastalhistory, fediphilosophy.org, newsie.social, writing.exchange,
urbanists.social, mastodon.gamedev.place, peoplemaking.games, photog.social, mastodon.education,
educator.social, mastodon.oeru.org\\
\bottomrule
\end{tabular}
\caption{List of field-specific servers.}
\label{tab:topic_specific_servers}
\end{table*}

\begin{figure*}[t]
  \centering
  \begin{subfigure}{0.48\textwidth}
    \centering
    \includegraphics[width=\linewidth]{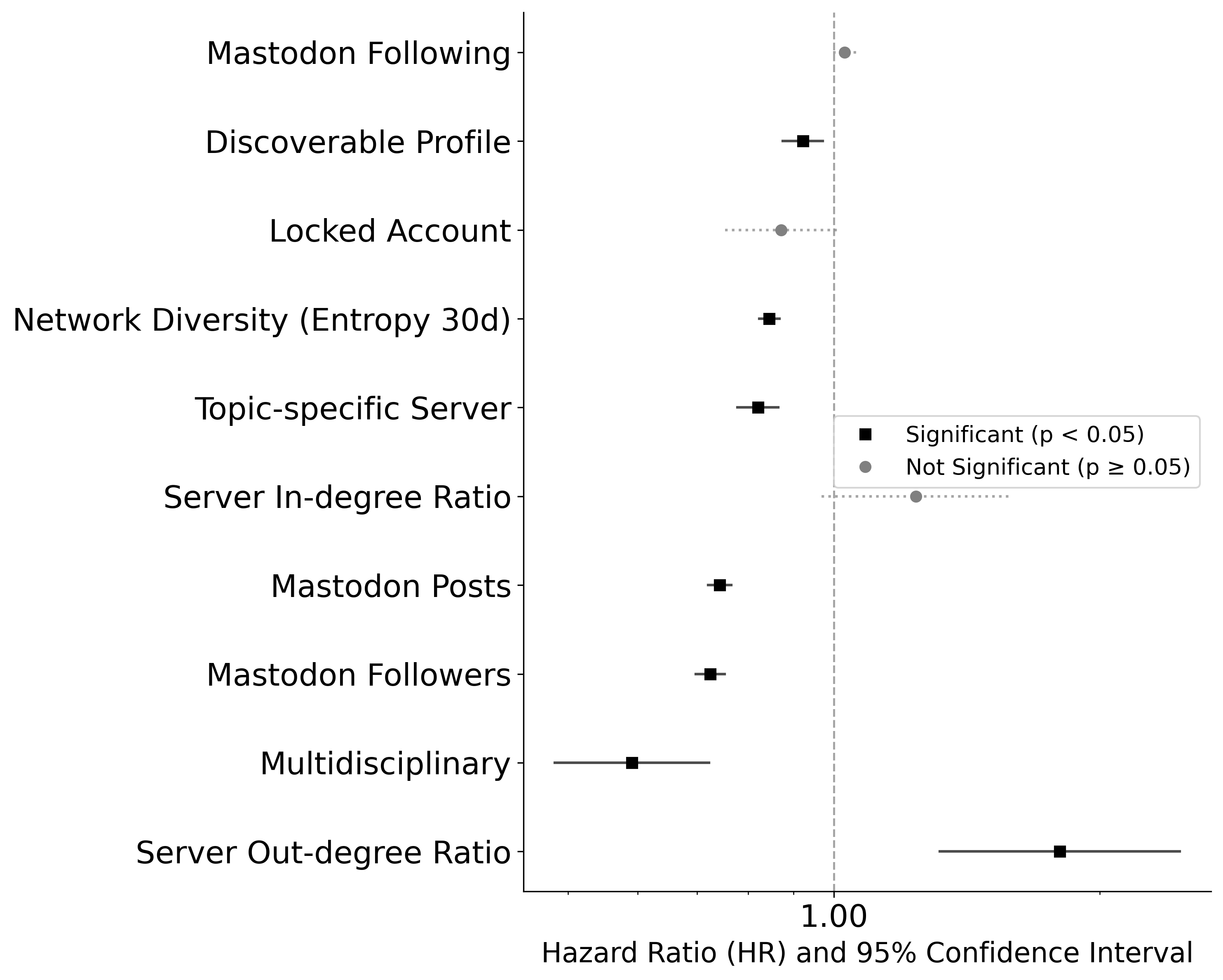}
    \caption{Primary Model: 30-Day Entropy Window (Concordance: 0.69)}
    \label{fig:mastodon_30d}
  \end{subfigure}
    \hfill
    \begin{subfigure}{0.48\textwidth}
        \centering
        \includegraphics[width=\linewidth]{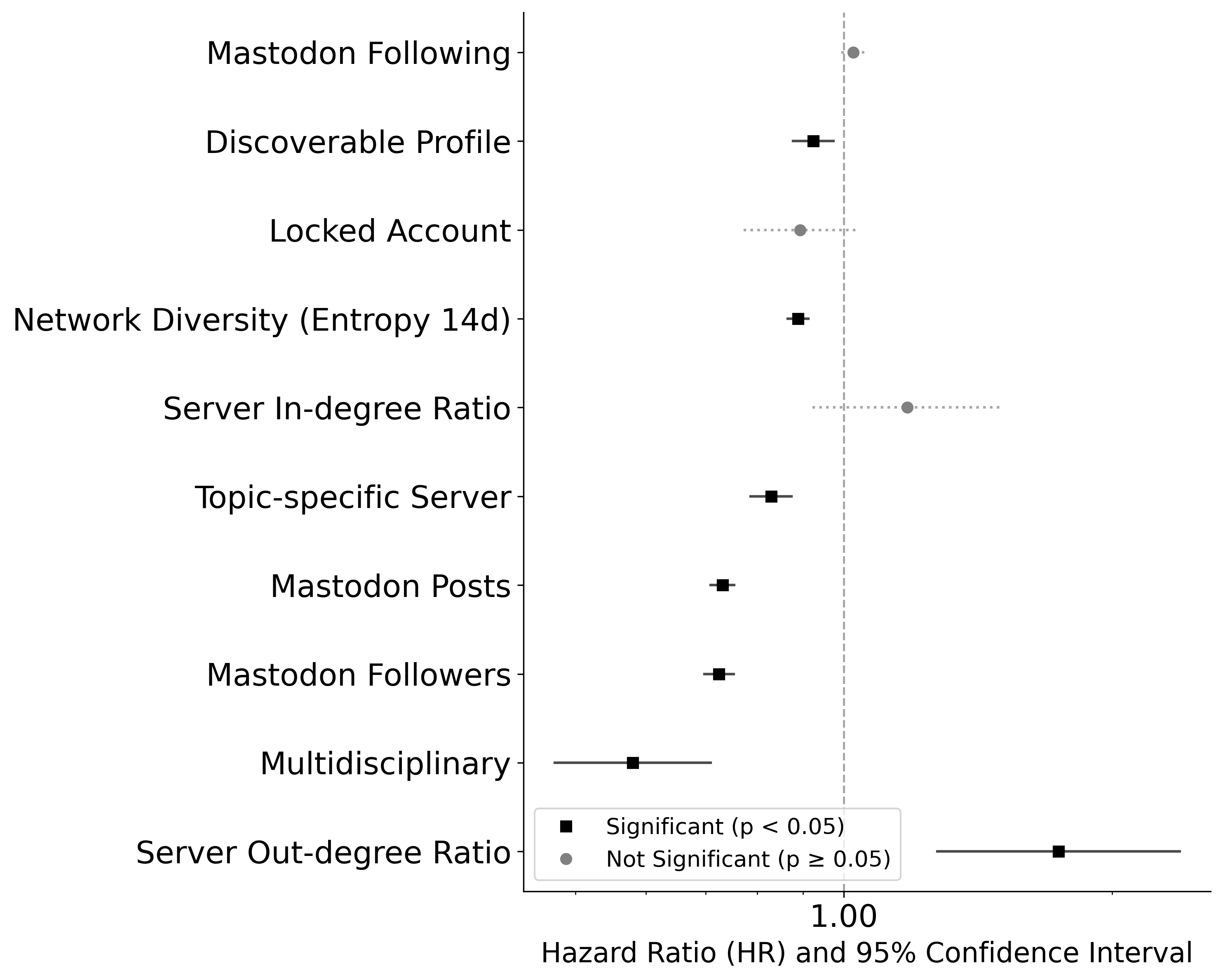}
        \caption{14-Day Entropy Window (Concordance: 0.68)}
        \label{fig:mastodon_14d}
    \end{subfigure}

    \begin{subfigure}{0.48\textwidth}
        \centering
        \includegraphics[width=\linewidth]{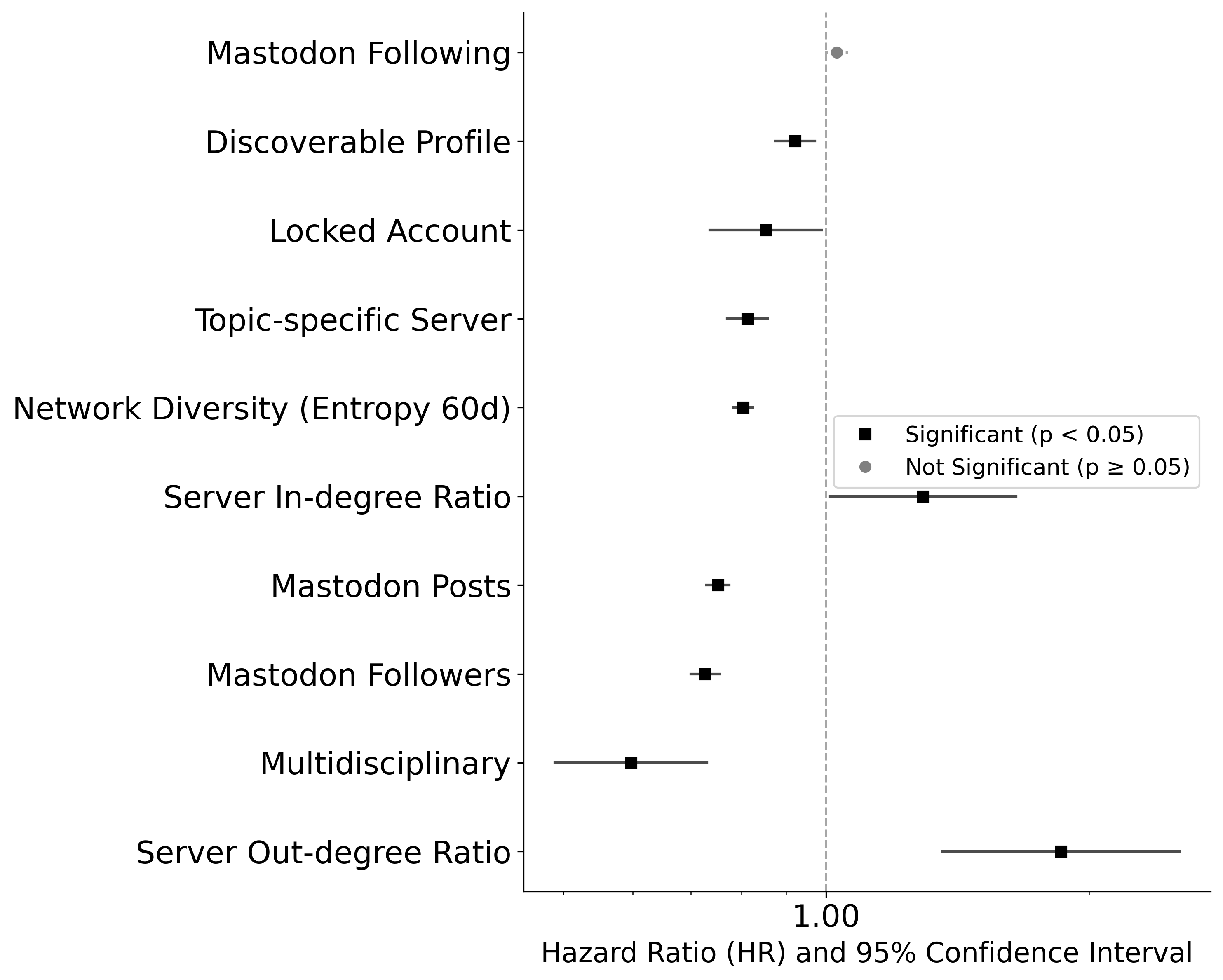}
        \caption{60-Day Entropy Window (Concordance: 0.69)}
        \label{fig:mastodon_60d}
    \end{subfigure}
    \hfill
    \begin{subfigure}{0.48\textwidth}
        \centering
        \includegraphics[width=\linewidth]{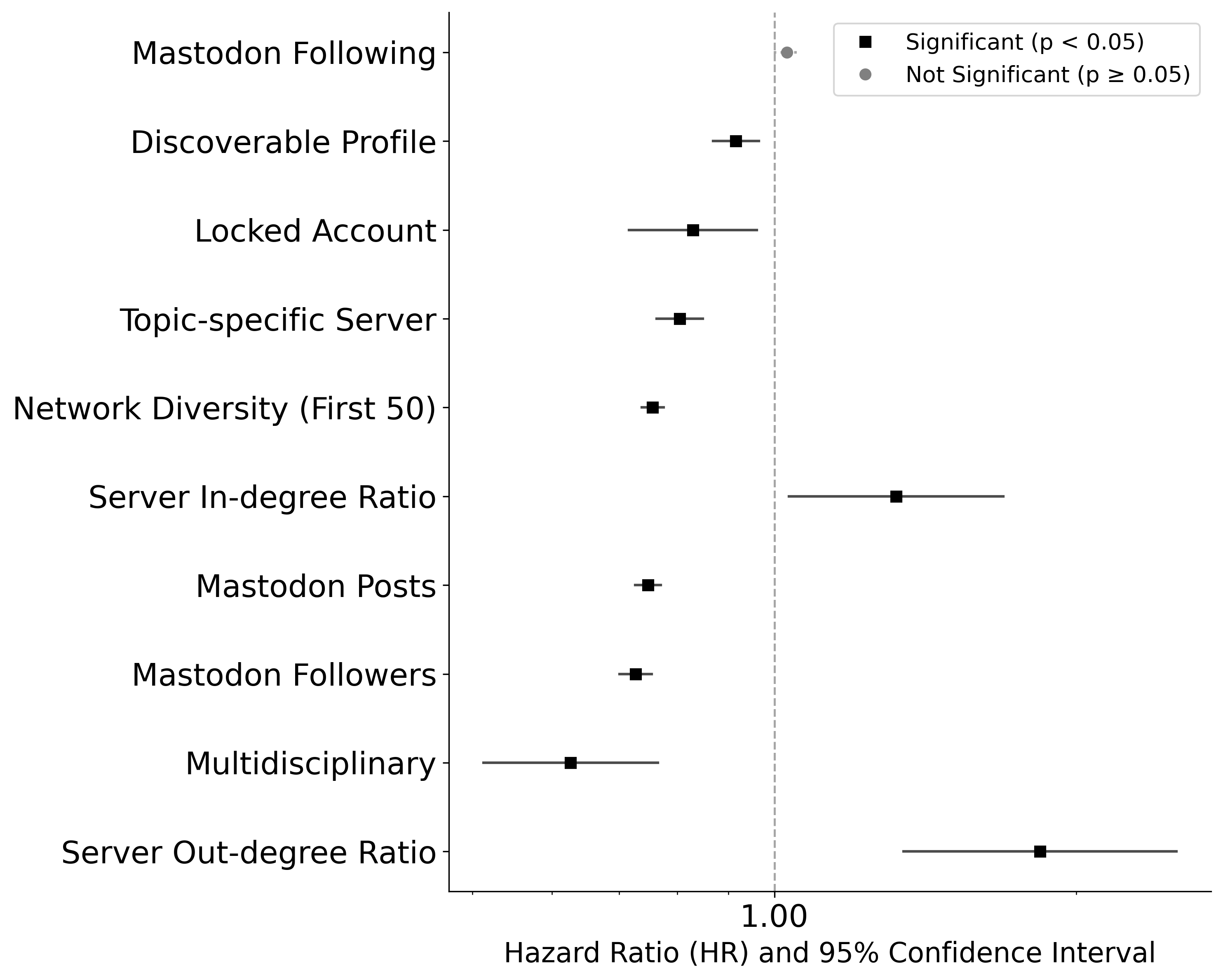}
        \caption{First 50 Interactions (Concordance: 0.70)}
        \label{fig:mastodon_first50}
    \end{subfigure}
    
    \caption{\textbf{Mastodon-Centric Model Sensitivity Analysis.} Forest plots displaying hazard ratios for predictors of inactivity across different specifications of the the federation interaction network metric. The protective effect of network diversity and multidisciplinary status remains robust across all time windows. Note that \textit{Locked Account} status becomes a significant risk factor only in the 60-day window ($p<0.05$), suggesting a cumulative effect.}
    \label{fig:model1_sensitivity}
\end{figure*}

\begin{figure*}[t]
  \centering
  \begin{subfigure}{0.48\textwidth}
    \centering
    \includegraphics[width=\linewidth]{Figs/forest_plot_full_30d.png}
    \caption{Primary Model: 30-Day Entropy Window (Concordance: 0.69)}
    \label{fig:mastodon_30d}
  \end{subfigure}
    \hfill
    \begin{subfigure}{0.48\textwidth}
        \centering
        \includegraphics[width=\linewidth]{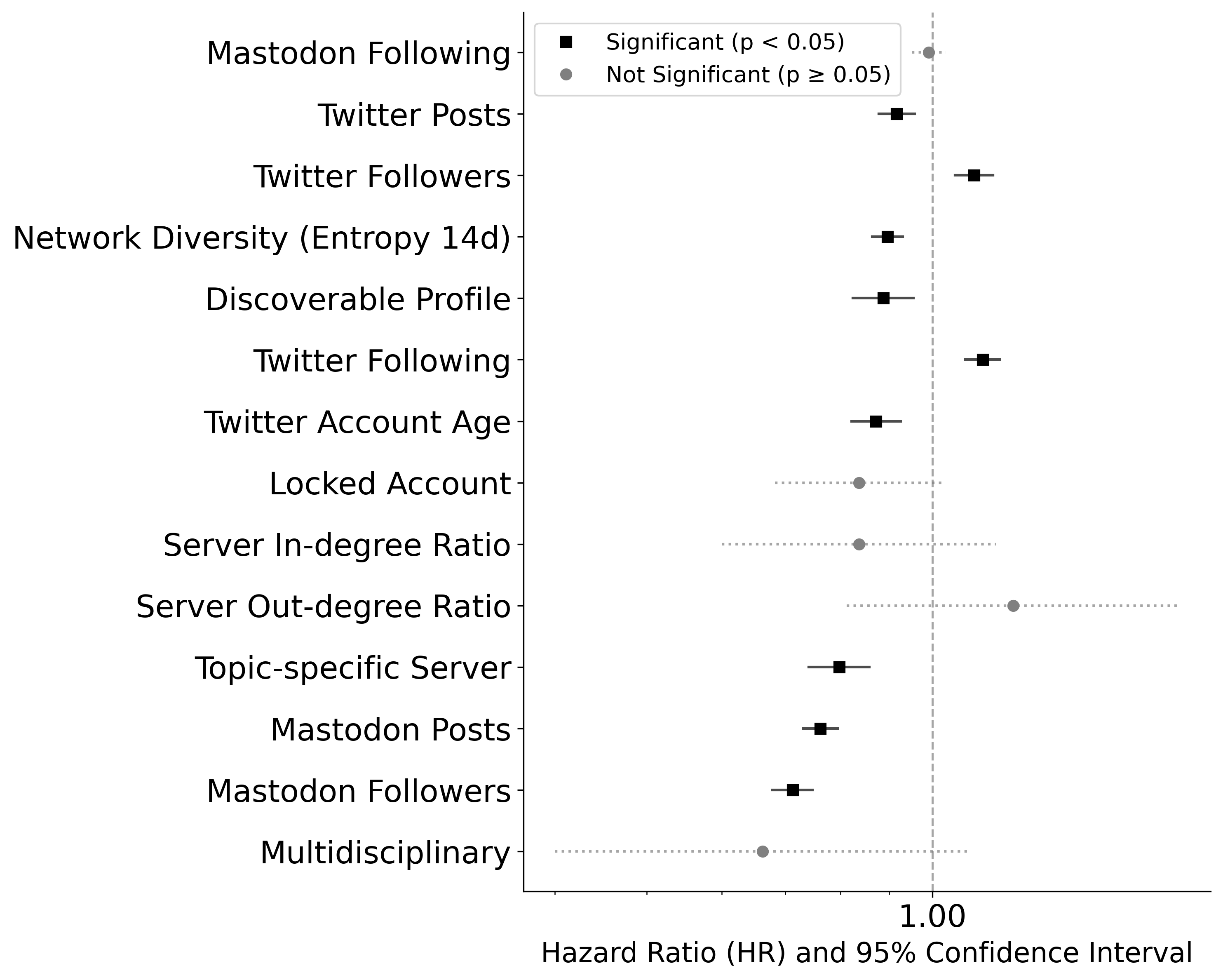}
        \caption{14-Day Entropy Window (Concordance: 0.70)}
        \label{fig:mastodon_14d}
    \end{subfigure}

    \begin{subfigure}{0.48\textwidth}
        \centering
        \includegraphics[width=\linewidth]{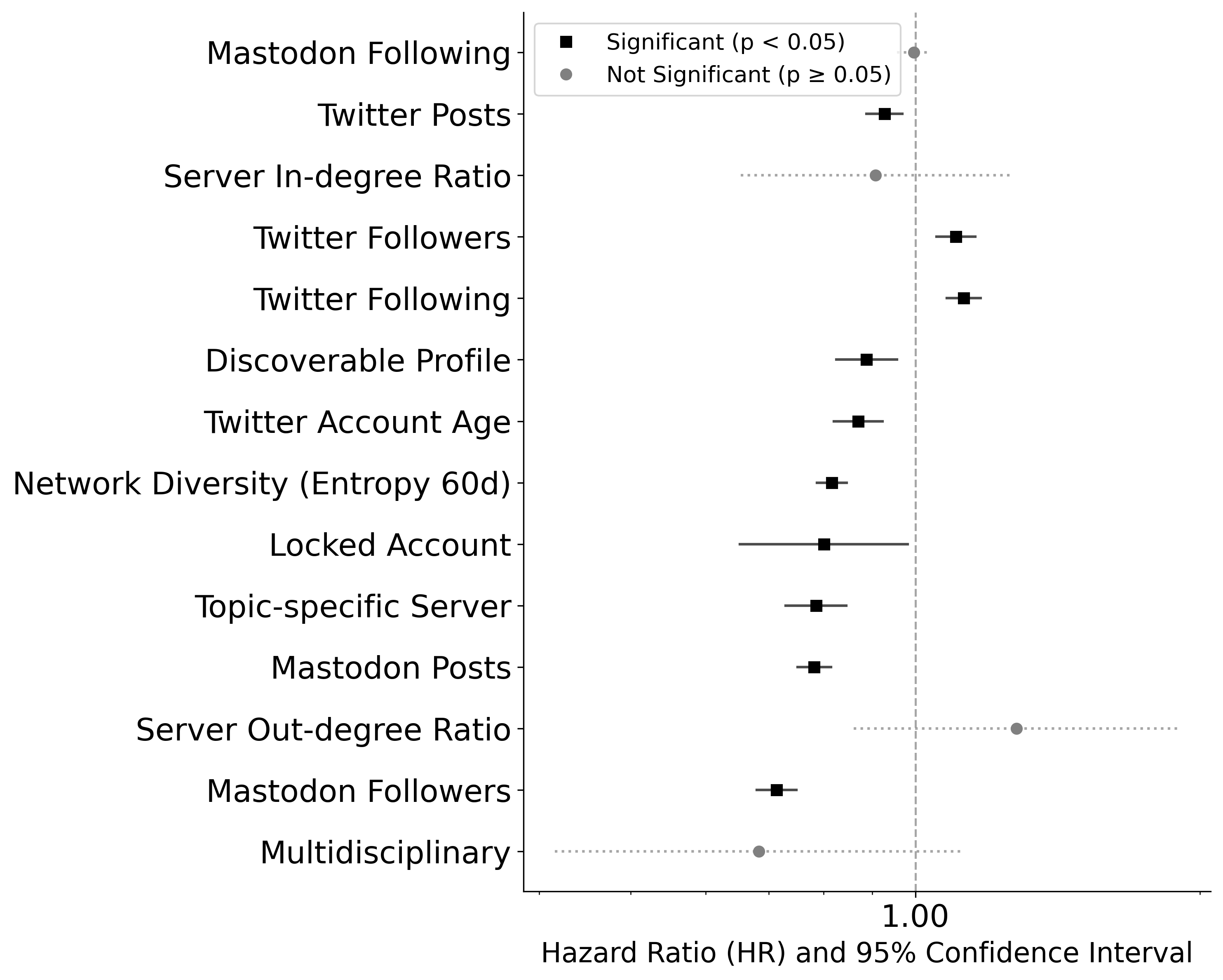}
        \caption{60-Day Entropy Window (Concordance: 0.69)}
        \label{fig:mastodon_60d}
    \end{subfigure}
    \hfill
    \begin{subfigure}{0.48\textwidth}
        \centering
        \includegraphics[width=\linewidth]{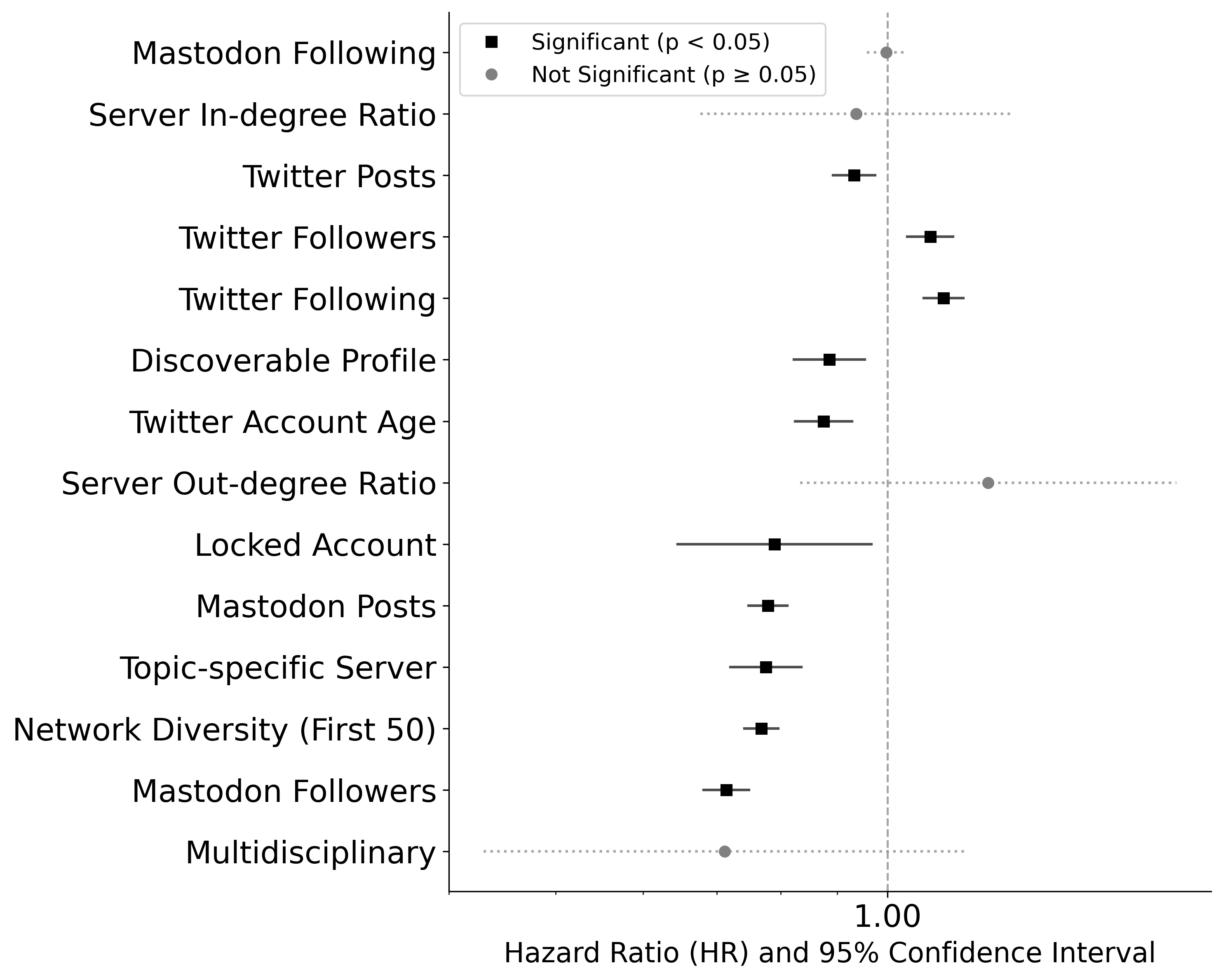}
        \caption{First 50 Interactions (Concordance: 0.70)}
        \label{fig:mastodon_first50}
    \end{subfigure}
    
    \caption{\textbf{Cross-Platform Model Sensitivity Analysis.} Forest plots for the subset of users with Twitter history. The plots confirm that Federated Interaction Diversity predicts retention independent of Twitter metadata. Unlike the Mastodon-centric model, the Multidisciplinary indicator is not statistically significant in this subset, while Twitter-derived metrics (Followers, Account Age) show consistent divergent associations with retention.}
    \label{fig:model2_sensitivity}
\end{figure*}

\end{document}